\documentclass[prb,twocolumn,showpacs,aps,superscriptaddress,floatfix]{revtex4}
\usepackage{amsmath}
\usepackage{amssymb}
\usepackage{bm}
\usepackage{graphicx}
\usepackage{color}
\usepackage{hyperref}
\usepackage{verbatim}
\graphicspath{{Figs/}}
\begin{document}
	\title{Bulk and surface spin conductivity in topological insulators with hexagonal warping}
	\author{R.S. Akzyanov}
	\affiliation{Moscow Institute of Physics and Technology, Dolgoprudny,
		Moscow Region, 141700 Russia}
	\affiliation{Institute for Theoretical and Applied Electrodynamics, Russian
		Academy of Sciences, Moscow, 125412 Russia}
	\affiliation{Dukhov Research Institute of Automatics, Moscow, 127055 Russia}
	\author{A.L. Rakhmanov}
	\affiliation{Moscow Institute of Physics and Technology, Dolgoprudny,
		Moscow Region, 141700 Russia}
	\affiliation{Institute for Theoretical and Applied Electrodynamics, Russian
		Academy of Sciences, Moscow, 125412 Russia}
	\affiliation{Dukhov Research Institute of Automatics, Moscow, 127055 Russia}
	
	\begin{abstract}
 We investigate the spin conductivity of topological insulators taking into account both the surface and quasi-two-dimensional bulk states. We apply a low-energy expansion of the Hamiltonian up to the third order in momentum and take into account the vertex corrections arising due to the short range disorder. Hexagonal warping gives rise to the additional anisotropic components in the spin conductivity tensor. Typically, isotropic part of the spin conductivity is larger than anisotropic one. The helical regime for the bulk states, in which the electrons in the Fermi level have the same projection of the spin on the direction of momentum, have been studied in a more detail. In this regime, a substantial increase of the spin conductivity contribution from the bulk states at the Fermi level is observed. We find that the bulk spin conductivity is insensitive to disorder if Rashba spin-orbit coupling is larger than disorder strength, otherwise, it is strongly suppressed. The contribution to the spin conductivity from the surface states is almost independent of the chemical potential, robust to disorder and its value is comparable to the spin conductivity contribution from the bulk states per layer. The obtained results are in agreement with experimental data.
\end{abstract}
	
	
	\maketitle
	\section{Introduction}

Topologically-protected surface states form a Dirac cone in the electronic spectrum of the topological insulators (TI)~\cite{RevModPhys.82.3045}. The electron dispersion near Dirac points is linear. However, a hexagonal warping of the Dirac cone arises when we take into account the next-order terms in the momentum expansion of corresponding Hamiltonian of the TIs with the hexagonal lattices, such as Bi$_2$Te$_3$~\cite{Fu2009} and Bi$_2$Se$_3$~\cite{Kuroda2010}. The hexagonal warping influences not only the surface states but also quasi-two-dimensional bulk states in these systems. Effects of the hexagonal warping on the electronic properties of the TI have been studied extensively~\cite{Wang2011,Pal2012,Siu2014,Repin2015}. In our recent paper~\cite{PhysRevB.97.075421}, we find that the presence of the hexagonal warping significantly affects the charge conductivity of the TI. In particular, it gives rise to the anisotropic anomalous in-plane magnetoresistance. Hexagonal warping also affects the quantum anomalous Hall effect and anomalous out-of-plane magnetoresistance.

A remarkable feature of the TIs is the existence of high spin conductivity in the absence of magnetic field, which is associated with an intrinsic spin Hall effect~\cite{Sinova2015}. This effect has been first predicted in Rashba spin-orbit coupled materials, such as GaAs~\cite{Murakami2003,Sinova2004}. However, the intrinsic spin Hall effect in such materials is weak due to short-range disorder (from a theoretical point of view, due to vertex corrections caused by this disorder)~\cite{Inoue2004,Raimondi2005}.

A change of direction of the magnetization in the magnetic material by a spin current is referred to as spin-transfer torque (STT)~\cite{Ralph2008}. The STT is closely related to the spin conductivity~\cite{Chen2015}. This effect can be used for the design of the fast and low dissipative magnetic memory~\cite{Wang2013}. Recent experiments reveal that STT in the TIs is by orders of magnitude larger than for any other material, which is a sign of a substantial spin conductivity in TIs~\cite{Mellnik2014,Fan2014,Fan2016,Han2017}. Experimental study of the STT in the TIs demonstrates some intriguing features. Both the in-plane and out-of-plane STT exist in the system, and the value of these effects is of the same order, which is unexpected from the spin-momentum locking argument~\cite{Mellnik2014}. Moreover, the sign of the spin conductivity may be different in different samples of the same material~\cite{Yang2016}. Spin conductivity in the TI is tuned by chemical potential and obeys a particle-hole asymmetry~\cite{Kondou2016,Fan2016}. Also, spin conductivity is suppressed in the bulk-insulating regime~\cite{Kondou2016}. It has been speculated that the large spin currents arise in the TIs due to the existence of the topologically-protected surface states~\cite{Shiomi2014,Wang2015a,Wang2016}. However, in the other papers it is complained that the spin conductivity in the TI mainly comes from the bulk states~\cite{Jamali2015,Han2017}.

In general, the spin conductivity includes both contributions from the states at the Fermi surface and from all filled states~\cite{Yang2006,Kodderitzsch2015}. While the contribution to the spin conductivity from the filled states can be calculated in a clean limit~\cite{Murakami2003,Sinova2004}, it is vital to treat the disorder correctly to describe the contribution from the states at Fermi level~\cite{Inoue2004,Raimondi2005}.

Unexpectedly small number of theoretical works are devoted to the spin conductivity in TIs. Recent DFT calculations of the contribution to the spin conductivity from the filled states show that quite large spin currents can exist in Bi$_x$Sb$_{1-x}$ and the value of the spin conductivity can be tuned by the chemical potential variation~\cite{Sahin2015}. In Ref.~\onlinecite{Peng2016}, the spin conductivity of the surface states in a thin film of TI with a cubic lattice has been studied neglecting the vertex corrections. The authors concluded that the dependence of the surface spin conductivity on the disorder and chemical potential is small. The spin conductivity in another Dirac material, graphene, attracted much more attention~\cite{Sinitsyn2006,Liu2015,Garcia2016}. Provided the spin-orbit interaction is induced in graphene, quite reasonable spin currents can be obtained in it. Recent calculations also show that large spin currents can be induced in Weyl semimetal, another Dirac material with large spin-orbit interaction~\cite{Sun2016}.

We study the spin conductivity of the surface and bulk states in the TI in a low energy approximation with taking into account the hexagonal warping. Both contributions from the filled states and from the states at the Fermi surface are considered. We apply the Kubo formalism accounting the vertex corrections to the velocity operators arising due to the short-range disorder. We show that the presence of the hexagonal warping leads to the additional anisotropic terms in the spin conductivity. We get that the spin conductivity is robust against disorder. The spin conductivity of the surface states is comparable with the spin conductivity of the bulk states per layer. The obtained results are consistent with the experimental data.

 The paper is organized as follows. In Section~\ref{sec_model} we analyze the Hamiltonian describing the surface and bulk states in the TI. In Section~\ref{sec_disorder} we introduce disorder and in Section~\ref{vertex} calculate the vertex corrections to the velocity operator. In Section~\ref{sec_fermi} we study the contribution to the bulk and surface spin conductivity from the states at the Fermi level. In Section~\ref{sec_top} we consider the contribution to the spin conductivity from the filled states. We estimate the values of the characteristic for TIs parameters in Section~\ref{sec_evaluation}. In Sec.~\ref{sec_discuss} we discuss the obtained results and compare them with the experiments and numerical calculations.

\section{Model}\label{sec_model}

Low energy surface and quasi-two-dimensional bulk states in the TI can be described by the Hamiltonian~\cite{PhysRevB.82.045122} ($\hbar=1$)
	\begin{eqnarray}\label{H0}
	\nonumber
	\hat{H}&=&r(k_x^2+k_y^2)+\mu+\alpha_{Rk}(k_x \sigma_y - k_y \sigma_x)\\
	&&+ \lambda k_x (k_x^2-3k_y^2)\sigma_z, \\
	\nonumber
	&&\alpha_{Rk}=\alpha_R[1+s(k_x^2+k_y^2)],
	\end{eqnarray}
where $\mathbf{\sigma}=(\sigma_x,\sigma_y,\sigma_z)$ are the Pauli matrices acting in spin space, $\mu$ is the chemical potential, $\alpha_R$ is the value of Rashba coupling (equal to the Fermi velocity for the surface states), $r=1/(2m)$ is the inverse mass term, $s$ characterizes the next order correction in momentum to $\alpha_R$, $k_x=k\cos \phi$ and $k_y=k\sin \phi$ are the in-plane momentum components, $\lambda$ is the hexagonal warping coefficient. The term in the Hamiltonian responsible for the hexagonal warping can be rewritten as $\lambda k_x (k_x^2-3k_y^2)=\lambda k^3 \cos 3\phi$ and the Hamiltonian is invariant under rotation on the angle $\phi=2\pi/3$. The spectrum of the Hamiltonian~\eqref{H0} is given by
	\begin{eqnarray}\label{spectrum}
	E_{\pm}=\mu +rk^2\pm
	\sqrt{\alpha_{Rk}^2k^2+\lambda^2 k^6\cos^2 3\phi}.
	\end{eqnarray}
If we measure the energy in terms of $\alpha_R^2/r$, then, the chemical potential, the next order correction to the spin-orbit coupling, and the hexagonal warping are conveniently characterized by the dimensionless values $r\mu /\alpha_R^2$, $s\alpha_R^2/r^2$, and $\lambda \alpha_R/r^2$, respectively.

\begin{figure}[t!]
\center
\includegraphics [width=7cm, height=7cm]{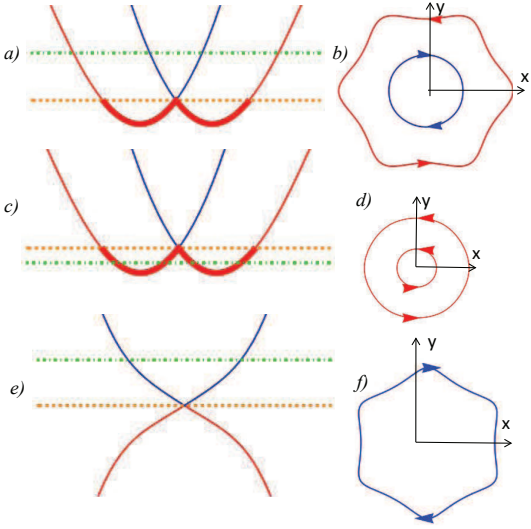}
\caption{Energy spectrum, Eq.~\eqref{spectrum}, and corresponding Fermi surface for different values of parameters. Spin direction of the states in the Fermi level is shown by arrows. Panels (a)--(d) illustrate the spectrum and the Fermi surface for the bulk states at $s=0$ and $\lambda\alpha_R/r^2=0.2$; $r\mu/\alpha_R^2=-2$ in (a) and (b), $r\mu/\alpha_R=0.2$ in (c) and (d). Bold lines in panels (a) and (c) indicate the helical regime. Panels (e) and (f) illustrate the spectrum and the Fermi surface for the surface states at $s\alpha_R^2/r^2=10$, $\lambda/s\alpha_R=0.2$, and $r\mu/\alpha_R=0.2$. Orange dashed lines indicate zero of the chemical potential and green dot-dashed lines show the chemical potential. }
		\label{spectra}
	\end{figure}

Energy spectrum~\eqref{spectrum} is shown in Fig.~\ref{spectra} for different set of parameters characteristic of the bulk, (a) and (c), and surface, (e), states. A key feature of the surface states in the TI is the existence of a robust Dirac cone, which is the case if $s\alpha_R^2/r^2$ is sufficiently large, Fig.~\ref{spectra}(e). Corresponding Fermi surface has a characteristic form of a snow-flake, Fig.~\ref{spectra}(f). The bulk states corresponds to smaller values of $s\alpha_R^2/r^2$, Figs.~\ref{spectra}(a) and (c). In the latter case, the spectrum has an appearance characteristic of a two-dimensional electron gas with bands splitted due to the Rashba spin-orbit interaction. Corresponding Fermi surfaces with two pockets are shown in Figs.~\ref{spectra}(b) and (d) for different values of the chemical potential. Note, that this model describes well ARPES data for the surface and bulk states in the TIs~\cite{Fu2009,Kuroda2010,PhysRevB.82.045122}.

We obtain from Eq.~\eqref{spectrum} that the robust Dirac cone exists when $s\alpha_R^2/r^2>1/3$. For $\alpha_R^2 (s+\lambda/\alpha_R)/r^2 < 1/4$, two spin split bands emerge in the system as it is expected for the Rashba spin-coupled electron gas. Therefore, we can formally to write down that
	\begin{eqnarray}\label{conditions}
	\alpha_R (s\alpha_R+\lambda)/r^2 &<& 1/4, \quad \textrm{bulk states}, \\ \nonumber
	\alpha_R^2 s/r^2 &>& 1/3, \quad \textrm{surface states}.
	\end{eqnarray}
In the case of the bulk states, the function $E_{-}(k)$ decreases at large $k$ and a proper momentum cut-off $k_{cut}$ must be introduced to avoid arising fake Fermi surface pockets at large momentum. We define cut-off momentum as $k_{cut}=r/(2s\alpha_R+2\lambda).$

We can calculate average spin projection of electrons as $\langle S_{\alpha} \rangle_{\pm} = \langle u_{\pm}| S_{\alpha} |u_{\pm}\rangle$, where $S_{\alpha}$ is the spin operator and $u_{\pm}$ are eigenfunctions corresponding to the bands $E_{\pm}$. The in-plane spin polarization component is schematically shown in Figs.~\ref{spectra}(b), (d), and (f). The calculated spin polarization lies in the $(x,y)$ plane if we neglect the hexagonal warping. We see that each band can be characterized by helicity, that is, the sign of the projection of the spin on the direction of momentum. The z-component of the spin polarization arises if we take into account that $\lambda\neq0$. If $\mu<0$ the bulk states have two splitted Fermi surfaces with different helicity, see panel (b). In the case of $\mu>0$ two Fermi surfaces have the same helicity, see panel (d), so, we call this regime for the bulk states as helical. Surface states are helical in any case, see panel (f).

In general, the spin conductivity can be presented as a sum of three terms~\cite{Yang2006,Kodderitzsch2015}
\begin{eqnarray}
\sigma_{{\alpha}\beta}^{\gamma}=\sigma_{{\alpha}\beta}^{\gamma I}+\sigma_{{\alpha}\beta}^{\gamma II}+\sigma_{{\alpha}\beta}^{\gamma III},
\end{eqnarray}
where the first two items correspond to a contribution from the states at the Fermi surface and the third one from the filled states. Here $\alpha$ and $\beta$ denote the in-plane coordinates $x$ and $y$, respectively, and $\gamma$ denotes the spin projection.

At zero temperature $\sigma_{{\alpha}\beta}^{\gamma I}$ and $\sigma_{{\alpha}\beta}^{\gamma II}$ can be written in the form~\cite{Inoue2004,Yang2006}
	\begin{eqnarray}\label{spin_surface}
	\sigma_{{\alpha}\beta}^{\gamma I}=\frac {e}{8\pi}\langle \textrm{Tr} [j_{\alpha}^{\gamma}\, G^+
\,V_{\beta}  \, G^- - j_{\alpha}^{\gamma}\, G^- \,V_{\beta}  \, G^+ ]\rangle,\\
	\label{spin_surface1}
	\sigma_{{\alpha}\beta}^{\gamma II}=-\frac {e}{8\pi}\langle \textrm{Tr} [j_{\alpha}^{\gamma}\, G^+ \,V_{\beta}  \, G^+ + j_{\alpha}^{\gamma}\, G^- \,V_{\beta}  \, G^- ]\rangle.
	\end{eqnarray}
	Here $j_{\alpha}^{\gamma}= \{\sigma_{\gamma}, v_{\alpha}\}/4$,  $v_{\alpha}= \partial H /\partial k_{\alpha}$ is the is the velocity operator, $V_{\alpha}$ is the velocity operator with vertex corrections, $\{\,,\}$ means the anticommutator, and $G^{\pm}$ are the retarded and advanced Green functions, which will be specified in the next section.
	
	The contribution to the spin conductivity from the filled states  is~\cite{Sinova2004,Sinitsyn2004}
	\begin{eqnarray}\label{sigmaIII}
	\nonumber
	\sigma_{{\alpha}\beta}^{\gamma III}=\frac {e}{4\pi } \sum\limits_{\mathbf{k},n\neq n'}(f_{n\mathbf{k}}-f_{n'\mathbf{k}})\times\\
	\frac{ \textrm{Im}\langle u_{n'\mathbf{k}}|j_{\alpha}^{\gamma}|u_{n\mathbf{k}}\rangle\langle u_{n\mathbf{k}}|v_{\beta}|u_{n'\mathbf{k}}\rangle}{\Gamma^2+(E_{n\mathbf{k}}-E_{n'\mathbf{k}})^2}.
	\end{eqnarray}
	Here $E_{n\mathbf{k}}$ is the energy of an electron in the $n$-th band with the momentum $\mathbf{k}$, $u_{n\mathbf{k}}$ is the corresponding Bloch vector, $\hat{H} u_{n\mathbf{k}} = E_{n\mathbf{k}} u_{n\mathbf{k}}$, $f_{n\mathbf{k}}$ is the Fermi distribution function corresponding to $E_{n\mathbf{k}}$ (which is the Heaviside step function in the considered case of zero temperature),  $\langle ... \rangle$ means impurity averaged, and $\Gamma$ is the disorder parameter or scattering rate. The latter will be also specified in the next section.

	\section{Disorder}\label{sec_disorder}

We will describe disorder by a potential $V_{\textrm{imp}}=u_0 \sum\limits_i \delta(\mathbf{r}-\mathbf{R}_j)$, where $\delta(\mathbf{r})$ is the Dirac delta function, $\mathbf{R}_j$ are positions of the randomly distributed point-like impurities with the local potential $u_0$ and concentration $n_i$. We assume that the disorder is Gaussian, that is, $\langle V_{\textrm{imp}} \rangle=0$ and $\langle V_{\textrm{imp}}(\mathbf{r}_1) V_{imp}(\mathbf{r}_2) \rangle=n_i u_0^2 \delta (\mathbf{r}_1-\mathbf{r}_2)$.

In the self-consistent Born approximation (SCBA), the impurity-averaged Green's functions can be calculated as
	\begin{eqnarray}\label{dyson}
	G^{\pm}=G_0^{\pm}+G_0^{\pm}\Sigma^{\pm} G^{\pm}
	\end{eqnarray}
where $G_0^{\pm}$ are bare Green's functions of the Hamiltonian~\eqref{H0}
		\begin{eqnarray}\label{green0}
		G^{\pm}_0\!\!=\!\!\frac{\mu\!+\!rk^2\!\pm\! i0\!-\!\alpha_{Rk}(k_x \sigma_y \!-\! k_y \sigma_x)\! - \! \lambda k^3\cos{3\phi}\,\sigma_z\!}{\left[\mu\!+\!rk^2
			\!\pm\! i0\right]^2\!-\!\alpha_{Rk}^2k^2\!-\left(\lambda k^3\cos{3\phi}\right)^2}
		\end{eqnarray}
	and $\Sigma^{\pm}$ is the self-energy, which is defined as
	\begin{eqnarray}
	\Sigma^{\pm}=  \langle V_{\textrm{imp}} G^{\pm} V_{\textrm{imp}}\rangle.
	\end{eqnarray}
	In the case under consideration, we can calculate the self-energy $\Sigma^{\pm}=\Sigma'\mp i\Gamma$ using an expression similar to that derived in Ref.~\onlinecite{Shon1998}
		\begin{eqnarray}\label{SelfEnergy}
		\!\!\Sigma^{\pm}\!\!=\!\!\frac{n_i u_0^2}{(2\pi)^2}\!\! \int\!\!\!\frac{(\mu\!+\! rk^2\!-\!\Sigma^{\pm})\,k dk\,d\phi }{(\mu\!+\!rk^2\!-\!\Sigma^{\pm})^2\!-\!\alpha_{Rk}^2k^2\!-\!\left(\lambda k^3\cos{3\phi}\right)^2}.\,
		\end{eqnarray}
The function under integral in Eq.~\eqref{SelfEnergy} decays as $k^3$ when $k\rightarrow\infty$. Thus, the value of this integral is determined by zeros of the denominator.

The value $\Gamma$ is usually referred to as a disorder parameter or scattering rate. It determines the analytical properties of the Green's functions $G^\pm$, while $\Sigma'$ is only a small correction to the chemical potential since we consider here only the case of small disorder. Thus, we can neglect the real part of the self-energy $\Sigma'$ with the exception of some singular point, which will be specified below. If we put $\Sigma'=0$ and have in mind that $\Gamma$ is small in the limit of small disorder, we derive from Eq.~\eqref{SelfEnergy} an explicit formula for the scattering rate
	\begin{equation}\label{Gamma1}
	\Gamma(\mu)=\frac {n_i u_0^2}{(2\pi)^2}\int\limits_0^{+\infty}\int\limits_0^{2\pi}\! k dk\,d\phi\,\,\textrm{Im} G_0^+.
	\end{equation}

\subsection{Bulk states}

First, we consider the bulk states. In the simplest case, when $\lambda$ and $s$ tends to zero, we can derive an explicit formula for the scattering rate in two opposite limits, $r\Gamma/\alpha_R^2 \ll 1$ and $r\Gamma/\alpha_R^2 \gg 1$. If the chemical potential $\mu$ is negative, we obtain from Eq.~\eqref{Gamma1} following Ref.~\onlinecite{Kato2007},
\begin{equation}\label{Gamma_0}
\Gamma(\mu<0)=\Gamma_0=\frac{n_iu_0^2}{4r}.
\end{equation}
This value is independent of the chemical potential and the strength of the spin-orbit interaction. It is convenient to introduce dimensionless disorder parameter $\gamma_b=n_iu_0^2/(4\alpha_R^2)$. In these notations $\Gamma_0=\alpha_R^2\gamma_b/r$.

Effects of the spin-orbit coupling are not smeared by the disorder if $\alpha_R^2/r \gg \Gamma_0$ or, equivalently, $\gamma_b \ll 1$. We will call further the spin-orbit coupling strong if condition $\gamma_b \ll 1$ is satisfied. Otherwise, $\gamma_b \gg 1$, the spin-orbit coupling is weak.

In the helical regime, $\mu>0$, the behavior of $\Gamma(\mu)$ depends on the system parameters. If the spin-orbit coupling is weak, $\gamma_b \gg 1$, we get that $\Gamma(\mu=0)=\Gamma_0/2$ and $\Gamma(\mu)$ rapidly decays to zero with an increase of $\mu$. In the opposite limit of strong  strong spin-orbit coupling, $\gamma_b \ll 1$, we found from Eq.~\eqref{Gamma1} that the scattering rate increases if  $0<\mu<\alpha_R^2/4r$:
\begin{eqnarray}
\Gamma(\mu)=\frac{\Gamma_0 \alpha_R}{\sqrt{\alpha_R^2-4\mu r}}.
\end{eqnarray}
When the chemical potential attains the singularity point, $\mu=\alpha_R^2/4r$, the Fermi level crosses the bottom of the energy bands $E_{\pm}$ if $s$ and $\lambda\rightarrow 0$ [see Eq.~\eqref{spectrum}]. If $\mu>\alpha_R^2/4r$, the Fermi level occurs in the energy gap. If we apply self-consistent Eq.~\eqref{SelfEnergy} we get that at $\mu=\alpha_R^2/4r$ the real part of the self-energy vanishes and
\begin{equation}
\Gamma(\mu=\alpha_R^2/4r)=\Gamma_{\textrm{max}}=\left(\frac{\alpha_R^2}{2r} \Gamma_0^2\right)^{1/3}.
\end{equation}
Thus, in the helical regime the scattering rate increases significantly if the spin-orbit coupling is strong, $\Gamma_{\textrm{max}}/\Gamma_0=\left(1/2\gamma\right)^{1/3}\gg 1$.

\begin{figure}[t!]
\center
\includegraphics [width=8.5cm, height=6cm]{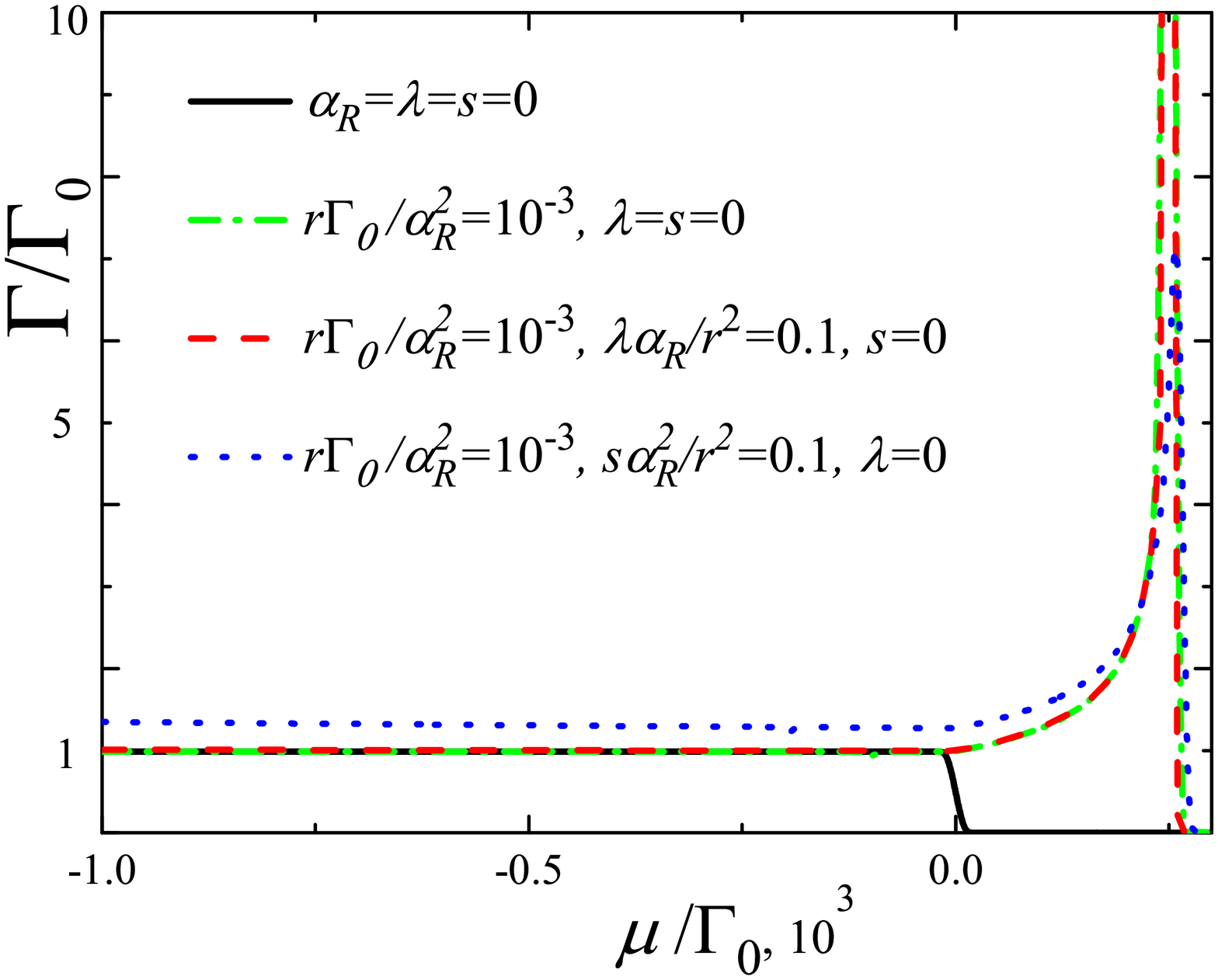}
\caption{Scattering rate $\Gamma$ for the bulk states as a function of the dimensionless chemical  potential $\mu/\Gamma_0$. Black line corresponds to the case, when the spin-orbit interaction and the hexagonal warping are absent, $\alpha_R, \lambda \rightarrow 0$. Green line corresponds to $r\Gamma_0/\alpha_R^2=0.001$, $\lambda=0$, and $s=0$; red line to $r\Gamma_0/\alpha_R^2=0.001$, $\lambda \alpha_R/r^2=0.1$, and $s=0$; blue line to $r\Gamma_0/\alpha_R^2=0.001$, $s \alpha_R^2/r^2=0.1$, and $\lambda=0$.}\label{gbulk}
	\end{figure}

In a more general case, the scattering rate $\Gamma(\mu)$ for the bulk states was calculated numerically using Eq.~\eqref{SelfEnergy}. The results are shown in Fig.~\ref{gbulk} for the case of the strong spin-orbit coupling characteristic of the TIs. As we can see from the figure, the higher order corrections to the spin-orbit coupling $s$ and the hexagonal warping $\lambda$ has a little impact on the value of the scattering rate in the case of the bulk states. In particular, a characteristic peak in $\Gamma(\mu)$ arises near the point $\mu=\alpha_R^2/4r$.

\subsection{Surface states}

For the surface states, we neglect a correction to the value of $\mu$ due to the real part of the self-energy in the limit of weak disorder, similar to the case of the bulk states. Thus, we can use Eq.~\eqref{Gamma1} to calculate $\Gamma$.

In the simplest case $r,\,\lambda,\,s\rightarrow 0$ we obtain a well-known result~\cite{Hu2008,Chiba2017},
\begin{eqnarray}\label{Gamma_2}
\Gamma(\mu)=\gamma_b |\mu|.
\end{eqnarray}
When the chemical potential crosses the Dirac point, $\mu=0$, we apply Eq.~\eqref{SelfEnergy} and find that the real part of the self-energy vanishes while imaginary part is exponentially small~\cite{Hu2008}
$
\Gamma(\mu=0)=\alpha_R \textrm{min} \{k_{cut},\sqrt{\alpha_R/\lambda},\sqrt{1/s} \} e^{-2/(\pi \gamma_b)}
$, where $k_{cut}$ is the momentum cut-off. That is, the scattering rate at the Dirac point is exponentially suppressed in the case of the strong spin-orbit coupling, $\gamma_b \ll 1$.

\begin{figure}[t!]
\center
\includegraphics [width=8.5cm, height=6cm]{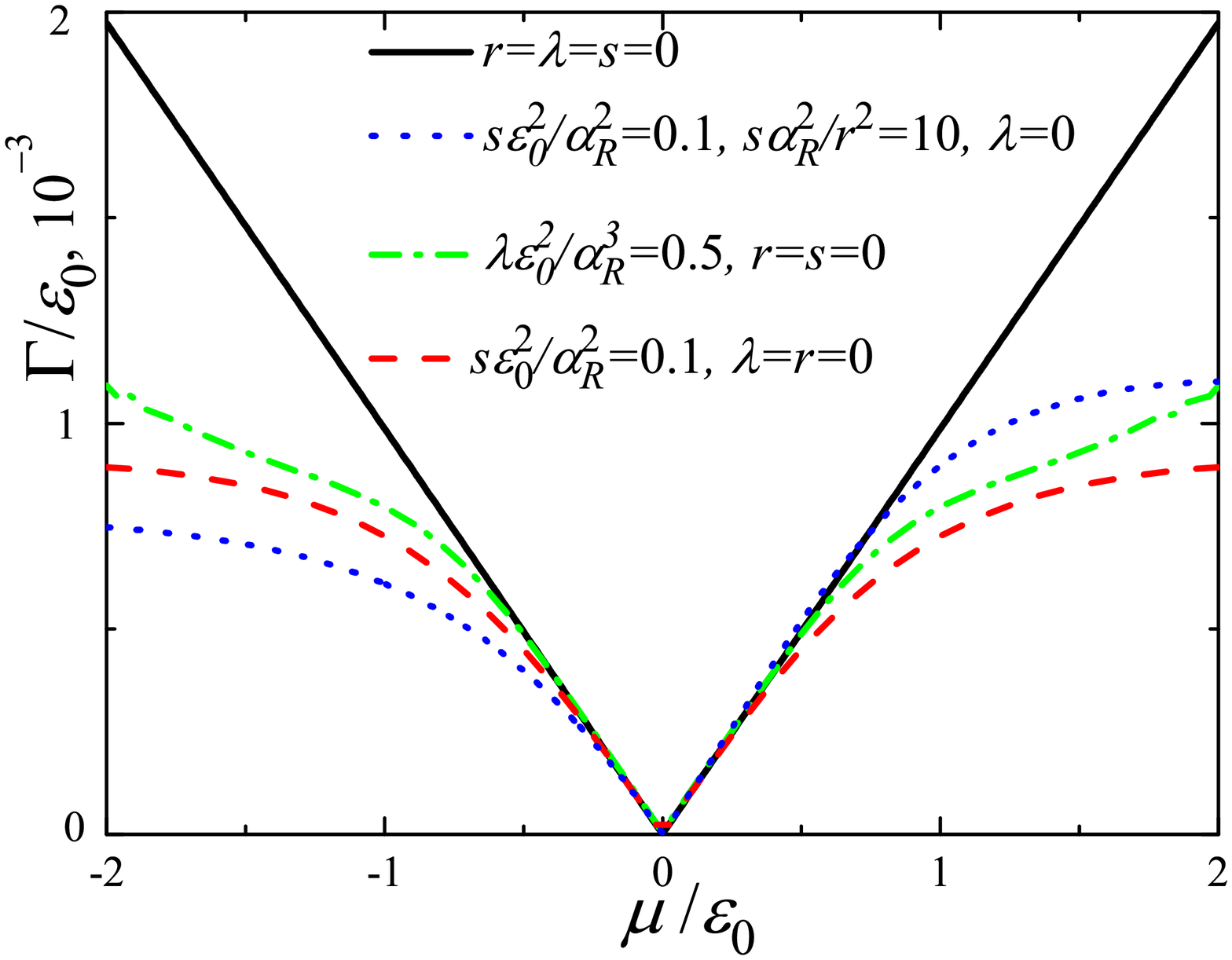}
\caption{Scattering rate $\Gamma$ for the surface states as a function of the chemical potential $\mu$ for $\gamma_b=0.001$. Black line corresponds to the case $r=s=\lambda=0$, green line to $\lambda \varepsilon_0^2/\alpha_R^3=0.5$, $s=r=0$, red line to $s \varepsilon_0^2/\alpha_R^2=0.1$, $\lambda=r=0$, blue line to $s \varepsilon_0^2/\alpha_R^2=0.1$, $\lambda=0$. Normalization parameter is chosen as $\varepsilon_0=\alpha_R k_{cut}/10$ where $k_{cut}$ is the cut-off momentum.}\label{gsurface}
	\end{figure}

In a more general case, the scattering rate for the surface states was calculated numerically with the help of Eq.~\eqref{SelfEnergy}. The dependence of $\Gamma(\mu)$ is shown in Fig.~\ref{gsurface}. We see that the scattering rate $\Gamma(\mu)$ is almost particle-hole symmetric since the spectrum of the surface states close to such a symmetry [see Fig.~\ref{spectra}(e)].

The self-energy is proportional to the identity matrix. This allows to obtain an explicit expression for the impurity averaged Green function. We can rewrite Eq.~\eqref{dyson} as $G^{\pm}=(1+\Sigma G_0^{\pm})^{-1} G_0^{\pm}$ or
		\begin{eqnarray}\label{greenF0}
\!\!\!\!G^{\pm}\!\!=\!\frac{\mu\!+\!rk^2\!\pm \!i\Gamma\!-\!\alpha_{Rk}(k_x \sigma_y \!-\! k_y \sigma_x)\! - \! \lambda k^3\cos{3\phi}\, \sigma_z\!}{(\mu+rk^2\pm i\Gamma)^2-\alpha_{Rk}^2k^2-\lambda^2 k^6\cos^2{3\phi}}\,.
		\end{eqnarray}
Therefore, the expression for $G^\pm$ is given by an equation similar to Eq.~\eqref{green0} for $G^\pm_0$, in which $\pm i0$ is replaced by $\pm i\Gamma$. We characterize disorder by a single value $\Gamma$ neglecting renormalization of the chemical potential.

\section{Vertex corrections}\label{vertex}

In the SCBA, following the approach described in Ref.~\onlinecite{Shon1998}, we can derive an equation for the vertex corrected velocity operator~\cite{Chiba2017}
\begin{eqnarray}\label{vert_corr}
V_{\alpha}(\mathbf{k})=v_{\alpha}(\mathbf{k}) + \frac{n_i u_0^2}{(2\pi)^2} \int G^+(\mathbf{k}) V_{\alpha}(\mathbf{k}) G^-(\mathbf{k})d^2\mathbf{k}.
\end{eqnarray}
We present here the derivation of $V_x$; results for $V_y$ can be obtianed just by the substitution $x(y)\rightarrow y(x)$.

It is easy to show that $n_i u_0^2\int G^+ v_x G^-d^2\mathbf{k}/(2\pi)^2=\zeta \sigma_y$ and $n_i u_0^2\int G^+ \sigma_y G^-d^2\mathbf{k}/(2\pi)^2=\kappa \sigma_y$, where $\zeta$ and $\kappa$ are scalars. In these notations we obtain from Eq.~\eqref{vert_corr} that
\begin{eqnarray}\label{V_FVC}
V_x=v_x+\left(\alpha_R^{VC}-\alpha_R\right)\sigma_y,\quad \alpha_R^{VC}=\alpha_R+\frac{\zeta}{1-\kappa}.
\end{eqnarray}

We begin our consideration with the bulk states and derive some analytical results in the simplest case, when $s$ and $\lambda$ are zero. Under such conditions and if $\mu<0$, the vertex corrected spin-orbit coupling is small, $\alpha_R^{VC}\ll \alpha_R$, either at weak, $\gamma_b\gg 1$, or strong, $\gamma_b \ll 1$, spin-orbit coupling in accordance with the results of previous works (see, e.g., Refs.~\onlinecite{Inoue2004,Raimondi2005}). However, in the case of strong spin-orbit coupling, when the chemical potential is positive and lies in the interval $0<\mu<\alpha_R^2/4r$, the vertex corrected $\alpha_R^{VC}$ is comparable to its bare value $\alpha_R$
	\begin{eqnarray}\label{vc_analytic}
	\frac{\alpha_R^{VC}}{\alpha_R}= \frac{\alpha_R-\sqrt{\alpha_R^2-4r\mu}}{\alpha_R+\sqrt{\alpha_R^2-4r\mu}}.
	\end{eqnarray}

\begin{figure}[t!]
\center
\includegraphics [width=8.5cm, height=6cm]{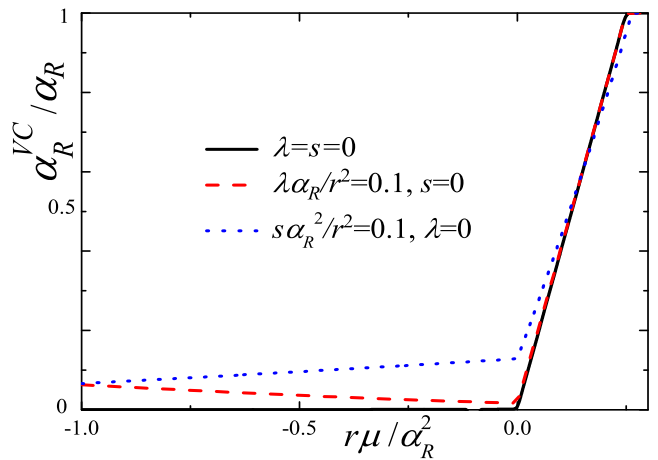}
\caption{Dependence of $\alpha_R^{VC}$ for the bulk states on the chemical potential, $\gamma_b=0.001$. Black line corresponds to $\lambda=s=0$, red line to $\lambda \alpha_R/r^2=0.1$ and $s=0$, blue line to $s\alpha_R/r^2=0.1$ and $\lambda=0$.}
\label{vertexbulk}
\end{figure}

Numerically calculated dependence of the vertex correction to the spin-orbit coupling for the bulk states on the chemical potential, $\alpha_R^{VC}(\mu)$, is shown in Fig.~\ref{vertexbulk} for a more general situation. As we can see, the higher order corrections to the spin-orbit coupling and the hexagonal warping significantly enhances $\alpha_R^{VC}$ in the region $\mu<0$. Nevertheless, its value is still much smaller than the bare value $\alpha_R$. When $\mu<0$ (in the helical regime), $\alpha_R^{VC}(\mu)$ is of the order of $\alpha_R$ and almost independent of $s$ and $\lambda$.

\begin{figure}[t!]
		\center
		\includegraphics [width=8.5cm, height=6cm]{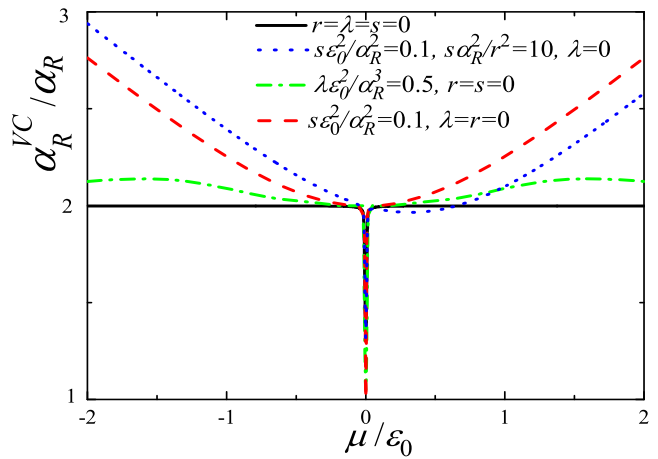}
		\caption{Dependence of the vertex correction $\alpha_R^{VC}$ on the chemical potential for the surface states, $\gamma_b=0.001$. Black line corresponds to the case $r=\lambda=s=0$, blue line to $s \varepsilon_0^2/\alpha_R^2=0.1$, $s \alpha_R^2/r^2=10$, and $\lambda=0$, green line to $\lambda \varepsilon_0^2/\alpha_R^3=0.5$ and $r=s=0$, red line to $s \varepsilon_0^2/\alpha_R^2=0.1$ and $r=\lambda=0$.}
		\label{vertexsurface}
	\end{figure}

For the surface states, in the case $r,\,s,\,\lambda=0$ and $\gamma_b \ll 1$, we get that away from the Dirac point, $\mu \gg \Gamma$, the vertex correction is $\alpha_R^{VC}=2\alpha_R$, while $\alpha_R^{VC}$ vanishes at $\mu=0$, as it have been obtained in Refs.~\onlinecite{Adroguer2012,Chiba2017}. Numerically calculated dependence of the vertex correction $\alpha_R^{VC}$ on the chemical potential is shown in Fig~\ref{vertexsurface} for the parameters characteristic of the surface states. The presence of the hexagonal warping slightly increases $\alpha_R^{VC}$, while the existence of the finite mass term $r$ leads to the particle-hole asymmetry. Taking into account correction to the spin-orbit coupling $s$ results in a significant increase of $\alpha_R^{VC}$ if the chemical potential is away from the Dirac point.

\section{Spin conductivity from the states at the Fermi surface}\label{sec_fermi}


Now we use the results obtained in the previous sections and Eqs.~\eqref{spin_surface} and \eqref{spin_surface1} to calculate the contribution to the spin conductivity due to the states at the Fermi surface. On this way, we obtained that in the considered approach the term $\sigma_{{\alpha}\beta}^{II\gamma}$ vanishes exactly. Thus, we should to compute only the term $\sigma_{{\alpha}\beta}^{I\gamma}$.

Isotropic tensor component $\sigma_{xy}^{Iz}=-\sigma_{yx}^{Iz}$ is the only term that persists in the system in the case of zero hexagonal warping. All other components are anisotropic and they are non-zero only if $\lambda\neq0$. The measured value of the spin conductivity depends on the mutual orientation of the current and the crystal axes. So, it is convenient to relate the spin conductivity tensor components in the crystal axes $(x,y)$ with that related to the current direction, $(\bar{x},\bar{y})$. New coordinates are obtained by anticlockwise rotation by the angle $\theta$ along the crystal axes. We assume that the current is directed along $\bar{x}$-axis. In this coordinates we have
\begin{eqnarray}\label{new_axes}
\nonumber
\sigma_{\bar{x}\bar{x}}^{I\bar{x}}&=&-\sigma_{\bar{y}\bar{y}}^{I\bar{y}}=-\sigma_{\bar{y}\bar{y}}^{I\bar{x}}=
	-\sigma_{\bar{y}\bar{x}}^{I\bar{y}}=\sigma_{xx}^{Ix}\cos 3\theta, \\
\sigma_{\bar{x}\bar{y}}^{I\bar{x}}&=&\sigma_{\bar{x}\bar{x}}^{I\bar{y}}=\sigma_{\bar{y}\bar{x}}^{I\bar{x}}=
	-\sigma_{\bar{y}\bar{y}}^{I\bar{y}}=-\sigma_{xx}^{Ix}\sin 3\theta,\\
\nonumber
 \sigma_{\bar{x}\bar{x}}^{Iz}&=&\sigma_{\bar{y}\bar{y}}^{Iz}=0.
\end{eqnarray}
Therefore, it is sufficient to calculate $\sigma_{xy}^{Iz}$ and $\sigma_{xx}^{Ix}$.

We derive from Eq.~\eqref{spin_surface}
		\begin{equation}\label{AAA}
		\begin{gathered}
		\sigma_{xy}^{Iz}=\sigma_0^z \int\!\! k\,dk\,d\phi\,\, \frac {2r\Gamma \alpha_{Rk} k^2 \left(\alpha_R^{VC}+\alpha_R s k^2\right)}{\pi^2 E_g(k,\phi)} \\
		\sigma_{xx}^{Ix}=\sigma_0^z \int\!\! k\,dk\,d\phi\,\, \frac {r\alpha_R\Gamma \lambda k^4(3+2sk^2)}{\pi^2 E_g(k,\phi)} \\
E_g\!=\!4\Gamma^2(\mu+rk^2)^2+\left(\Gamma^2-E_-E_+\right)^2,
\end{gathered}
		\end{equation}
where $\sigma_{0}^{z}=e/(8\pi)$ is the spin conductivity quanta and $E_{\pm}$ is given by Eq.~\eqref{spectrum}. As we can see, the isotropic spin conductivity component $\sigma_{xy}^{Iz}$ is proportional to $\alpha_R^{VC}$ if we neglect the higher-order correction to the spin-orbit coupling. This value increases significantly in the helical state $\mu>0$. The anisotropic component $\sigma_{xx}^{Ix}$ is proportional to the hexagonal warping strength $\lambda$, the vertex corrections does not affect it.

	\begin{figure}[t!]
		\center
		\includegraphics [width=8.5cm, height=6cm]{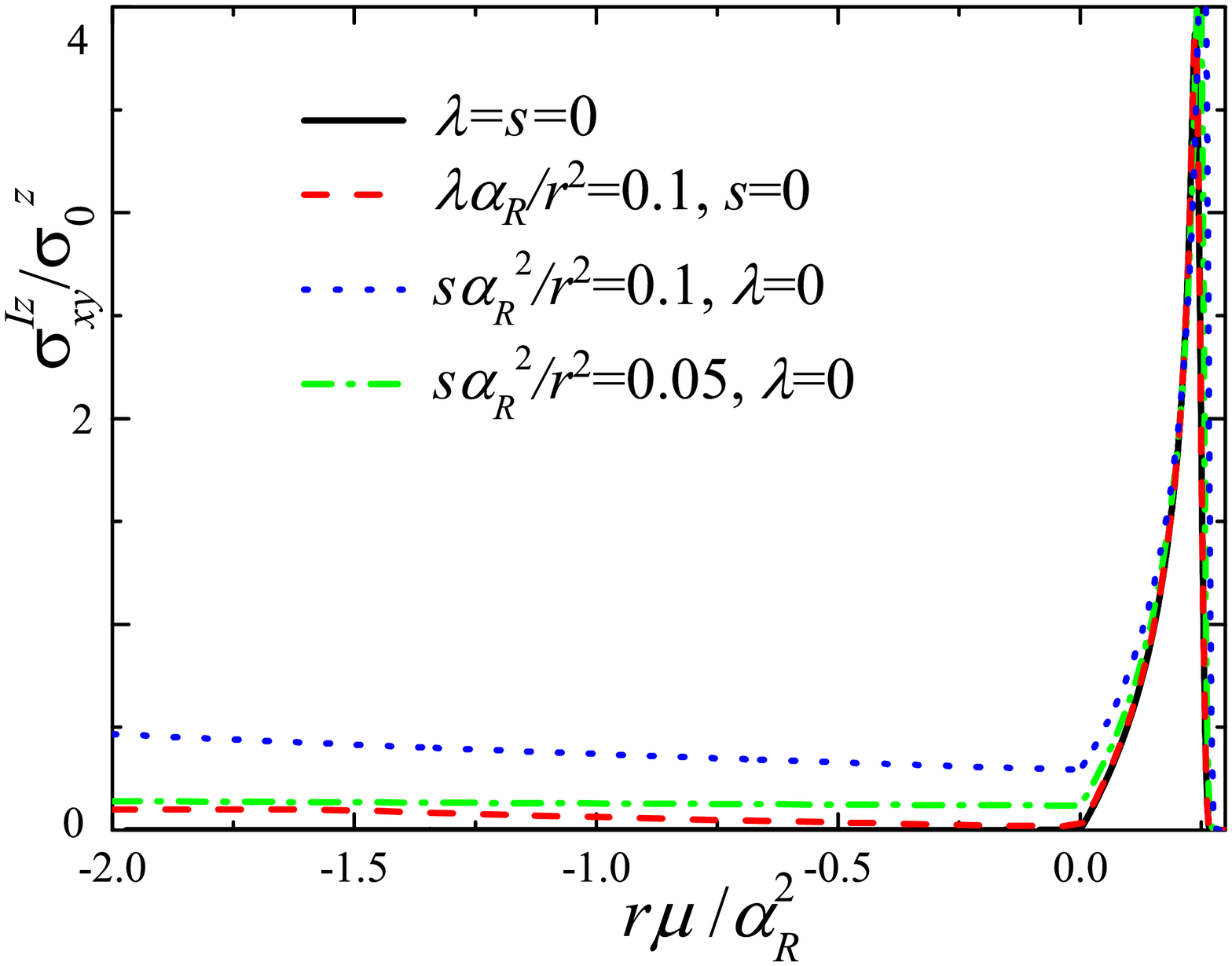}
		\includegraphics [width=8.5cm, height=6cm]{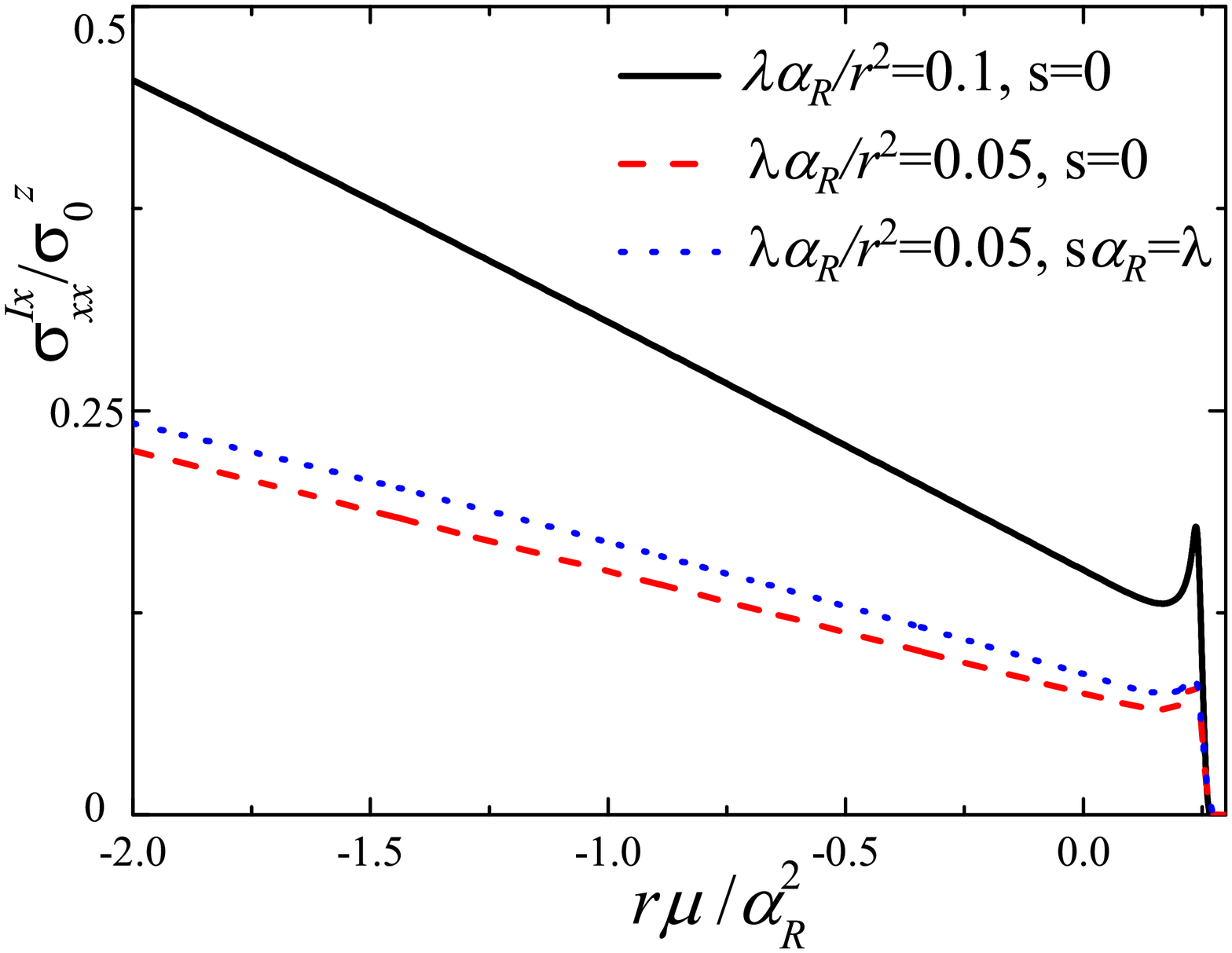}
		\caption{Isotropic spin conductivity $\sigma_{xy}^{Iz}$ (top panel) and anisotropic spin conductivity $\sigma_{xx}^{Ix}$ (bottom panel) for the bulk states as a function of the chemical potential $\mu$, $\gamma_b=0.01$. In the top panel black line corresponds to $s=\lambda=0$, red line to $\lambda \alpha_R/r^2=0.1$ and $s=0$, blue line to $s\alpha_R^2/r^2=0.1$ and $\lambda=0$, green line to $s\alpha_R^2/r^2=0.05$ and $\lambda=0$. In bottom panel black line corresponds to $\lambda \alpha_R/r^2=0.1$ and $s=0$, red to $\lambda \alpha_R/r^2=0.05$ and $s=0$, blue line to $\lambda \alpha_R/r^2=0.05$ and $s=\lambda/\alpha_R$ }\label{s1xyzbulku}
	\end{figure}

First, we calculate the contribution to the spin conductivity from the bulk states. The results are shown in Fig.~~\ref{s1xyzbulku}. As we can see from the top panel in Fig.~\ref{s1xyzbulku}, the isotropic spin conductivity component $\sigma_{xy}^{Iz}$ is suppressed when the chemical potential is negative. It occurs since the vertex correction $\alpha_R^{VC}$ is small in this region of  $\mu$. However, when $0<\mu<\alpha_R^2/4r$, the value of $\alpha_R^{VC}$ is comparable to $\alpha_R$ and the value of $\sigma_{xy}^{Iz}$ increases significantly. Note that $s$ and $\lambda$ produces a weak effect on $\sigma_{xy}^{Iz}$ in this range of $\mu$. If $\mu>\alpha_R^2/(4r)$, the spin conductivity vanishes since the density of states on the Fermi disappears.

The results for the anisotropic component of the spin conductivity $\sigma_{xx}^{Ix}$ are presented in the bottom panel of Fig.~\ref{s1xyzbulku}. This value decreases almost linearly with an increase of the chemical potential if $\mu<0$ and, when $\mu>0$, it has a small peak near $\mu=\alpha_R^2/4r$. The value $\sigma_{xx}^{Ix}$ also demonstrate almost a linear growth with an increase of the coefficients $\lambda$ and $s$.

	\begin{figure}[t!]
		\center
		\includegraphics [width=8.5cm, height=6cm]{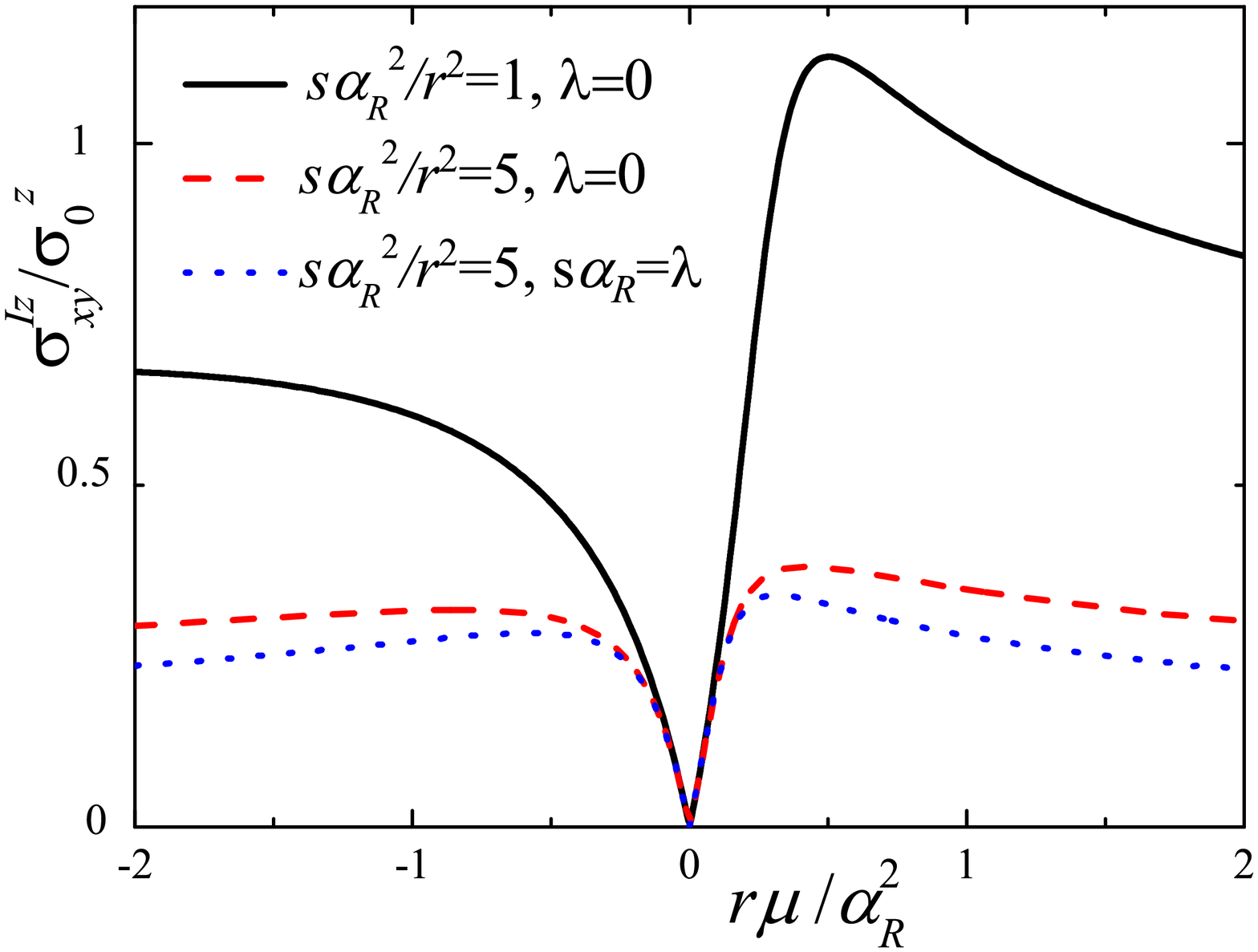}
		\includegraphics [width=8.5cm, height=6cm]{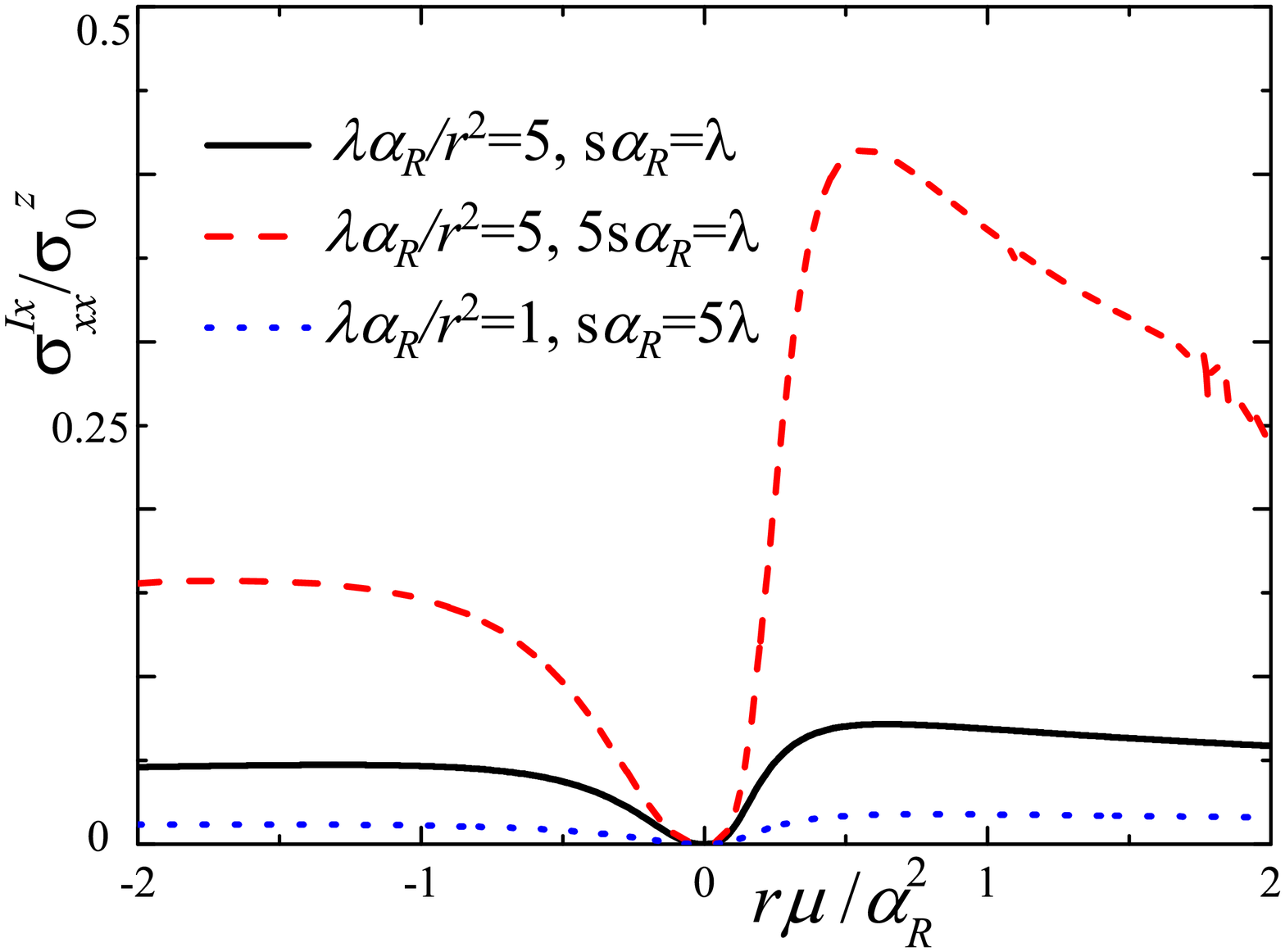}
		\caption{Isotropic spin conductivity $\sigma_{xy}^{Iz}$ (top panel) and anisotropic spin conductivity $\sigma_{xx}^{Ix}$ (bottom panel) for the surface states as a function of the chemical potential, $\gamma_b=0.01$. In top panel black line corresponds to $s\alpha_R^2/r^2=1$ and $\lambda=0$, red line to $s\alpha_R^2/r^2=5$ and $\lambda=0$, blue line to  $s\alpha_R^2/r^2=5$ and $\lambda=s\alpha_R$. In bottom panel black line corresponds to $s\alpha_R^2/r^2=5$ and $\lambda=s\alpha_R$, red line to $s\alpha_R^2/r^2=1$ and $\lambda=5s\alpha_R$, blue line to $s\alpha_R^2/r^2=5$ and $\lambda=s\alpha_R/5$. }\label{sxyzsurfaceu}
	\end{figure}

The dependencies of the isotropic and anisotropic components of the spin conductivity on $\mu$ for the surface states are shown in Fig.~\ref{sxyzsurfaceu}. Both these values have minima at the Dirac point $\mu=0$ and particle-hole asymmetry that is smaller for larger $s$. Note that $\sigma_{xy}^{Ix}$ and $\sigma_{xx}^{Ix}$ for the surface states decreases with an increase of the next order correction to the spin-orbit coupling coefficient $s$.

\begin{figure}[t!]
\center
\includegraphics [width=8.5cm, height=6cm]{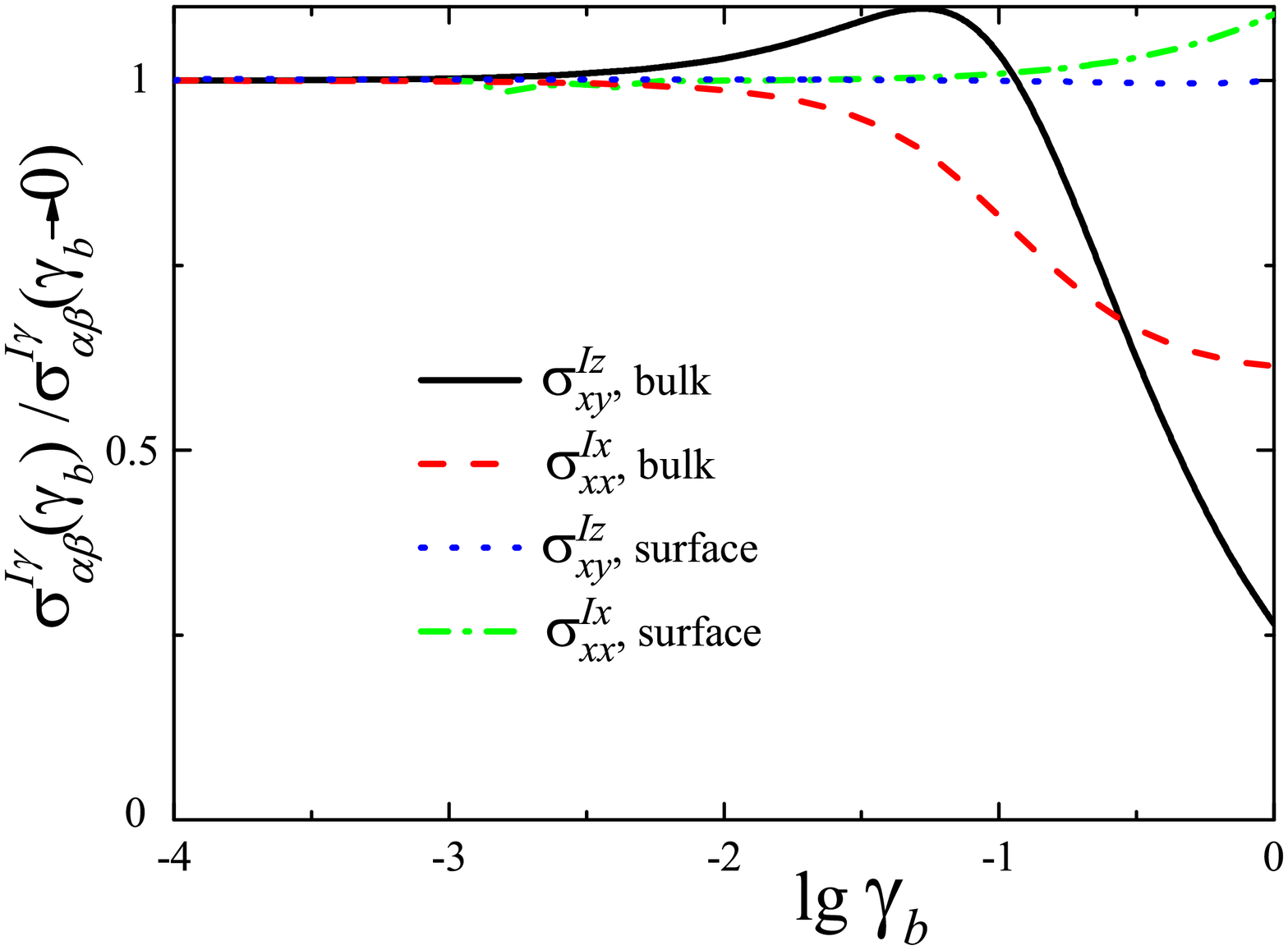}
\caption{Contribution to the spin conductivity from the states at the Fermi level, $\sigma_{\alpha \beta}^{I\gamma}$, for the bulk and surface states as a function of disorder strength $\gamma_b$. For all curves $\mu=-\alpha_R^2/r$. Black line corresponds to the contribution from the bulk states in $\sigma_{xy}^{Iz}$ for $s$, and $\lambda=0$, red line to the contribution from the bulk states in $\sigma_{xx}^{Ix} $ for $\lambda \alpha_R/r^2=0.1$ and $s=0$, blue line is the contribution from the surface states in $\sigma_{xy}^{Iz}$ for $s\alpha_R^2/r^2=5$ and $\lambda=0$, green line is the contribution from the surface states in $\sigma_{xx}^{Ix}$ for $s\alpha_R^2/r^2=5$ and $\lambda =s\alpha_R$.}\label{fermig}
\end{figure}

The effect of disorder on the spin conductivity $\sigma_{\alpha \beta}^{I\gamma}$ is illustrated in Fig.~\ref{fermig}. Both surface and bulk conductivities are robust against disorder in the weak scattering limit, $\gamma_b \ll 1$. Moreover, the topologically protected surface terms are robust even in the case of higher disorder, $\gamma_b \sim 1$, while the components of the bulk conductivity decrease significantly in this limit. However, the SCBA is not correct for a strong disorder $\gamma_b \sim 1$ and more advanced techniques are required to study the robustness of the spin conductivity of the surface states in such regime.

\section{Spin conductivity from the filled states}\label{sec_top}

Here we calculate the contribution to the spin conductivity from the filled states using Eq.~\eqref{sigmaIII} and the obtained above results for the disorder parameter $\Gamma$. Note that the vertex corrections do not affect this part of the spin conductivity. Similar to the spin conductivity from the states at the Fermi surface, the spin conductivity from the filled states has an isotropic component $\sigma_{xy}^{IIIz}$ and anisotropic one $\sigma_{xx}^{IIIx}$. The isotropic component, $\sigma_{xy}^{IIIz}=-\sigma_{yx}^{IIIz}$, is the only term that persists in the system in the absence of the hexagonal warping. The anisotropic components are non-zero only if $\lambda\neq0$. In the rotated coordinates $(\bar{x},\bar{y})$, the components of tensor $\sigma_{\alpha\beta}^{III\gamma}$ transform similar to $\sigma_{\alpha\beta}^{I\gamma}$, see Eq.~\eqref{new_axes}.

We obtain by means of Eq.~\eqref{sigmaIII}
\begin{equation}
		\begin{gathered}\label{BBB}
		\sigma_{xy}^{IIIz}\!=\!\sigma_0^z \int \left[\Theta(E_1)-\Theta(E_2)\right]k\,dk\,d\phi \frac {2rk\alpha_{Rk}^2}{\pi E_s} \\
		\sigma_{xx}^{IIIx}\!=\!\sigma_0^z \int\left[\Theta(E_1)-\Theta(E_2)\right] k\,dk\,d\phi \frac {rk^4\alpha_R\lambda (3+2k^2s)}{\pi E_s} \\
		E_s\!=\!\sqrt{\alpha_{Rk}^2+\lambda^2 k^6 \cos^2{3\phi}}\,\left[4\alpha_{Rk}^2k^2+\lambda^2k^6\cos^2{3\phi}
		+\Gamma^2\right],
\end{gathered}
\end{equation}
where $\Theta(x)$ is the Heaviside step function.

\begin{figure}[t!]
		\center
		\includegraphics [width=8.5cm, height=6cm]{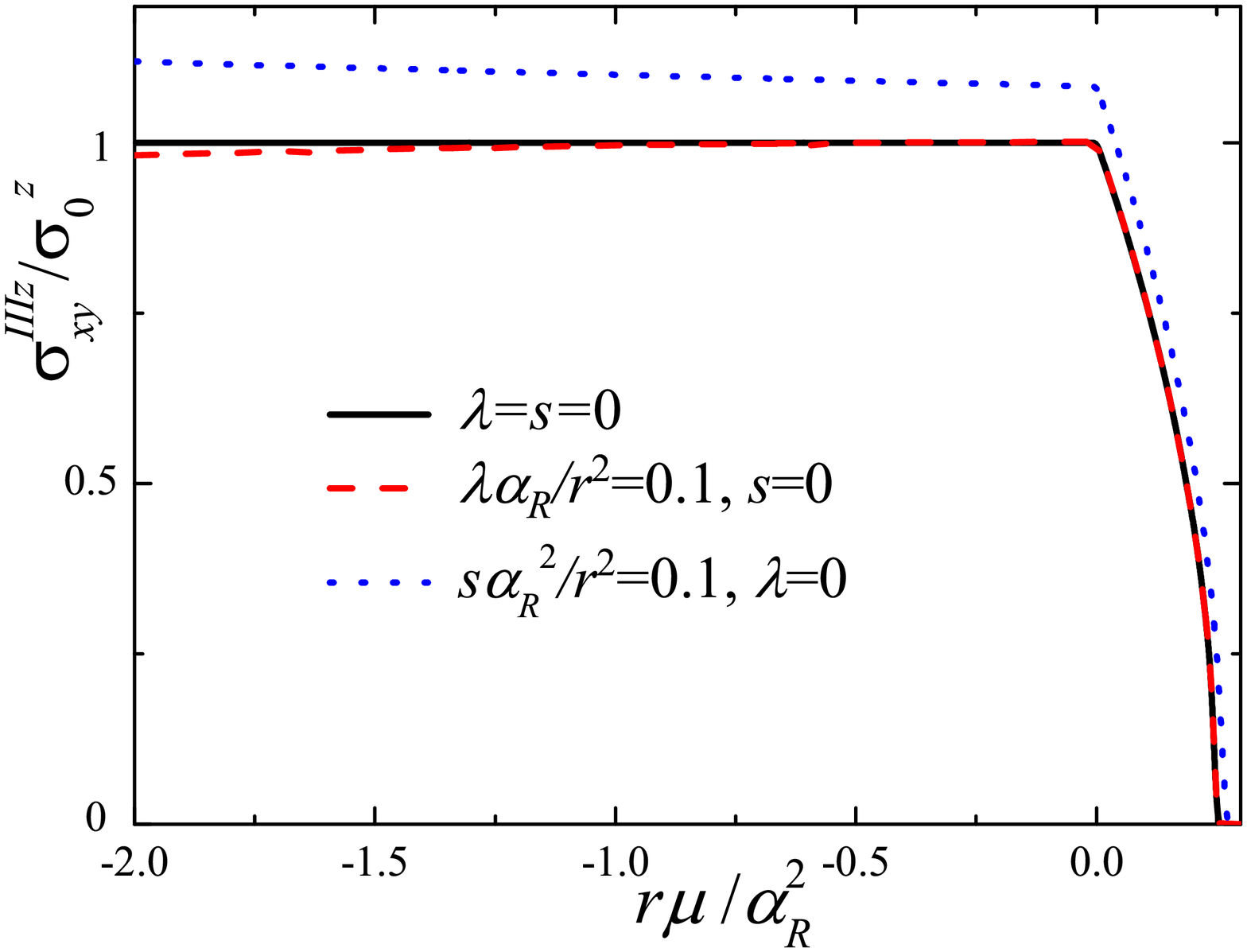}
				\includegraphics [width=8.5cm, height=6cm]{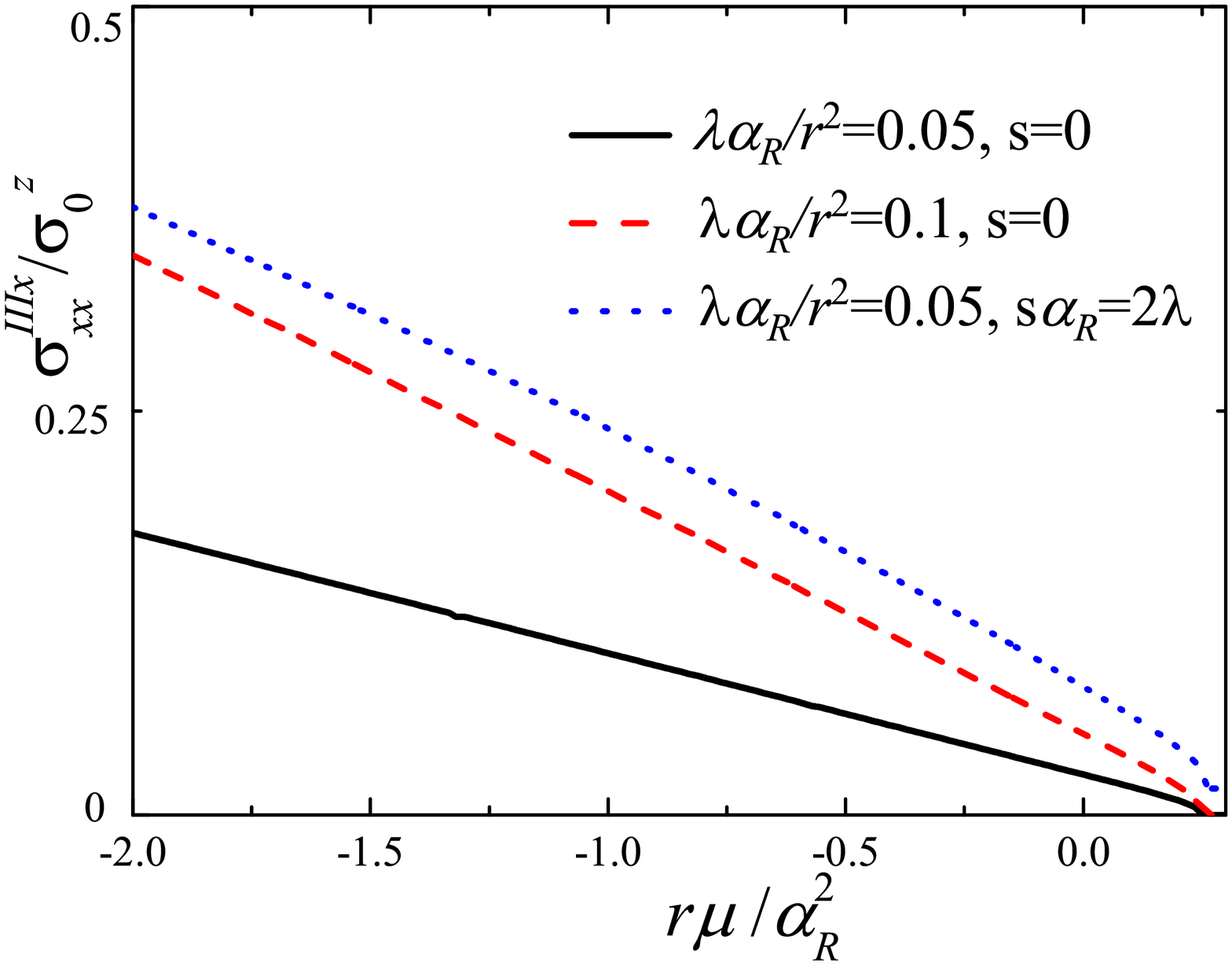}
		\caption{Isotropic spin conductivity $\sigma_{xy}^{IIIz}$ (top panel) and anisotropic spin conductivity $\sigma_{xx}^{IIIx}$ (bottom panel) for the bulk states as a function of the chemical potential in the clean limit, $\Gamma=0$. In top panel black line corresponds to $\lambda$ and $s=0$, red line to $\lambda \alpha_R/r^2=0.1$ and $s=0$, blue line to $s \alpha_R^2/r^2=0.1$ and $\lambda=0$. In bottom panel black line corresponds to $\lambda \alpha_R/r^2=0.05$ and $s=0$, red line to $\lambda \alpha_R/r^2=0.1$ and $s=0$, blue line to $\lambda \alpha_R/r^2=0.1$, $s\alpha_R=2\lambda$.}\label{sxyzbulktopu}
	\end{figure}

We start with the bulk spin conductivity. In the clean limit, $\Gamma=0$, and zero third order corrections, $\lambda=0$ and $s=0$, we get that  $\sigma_{xy}^{IIIz}=\sigma_0^z\sqrt{1-4\mu r/\alpha_R^2}$ if $\mu>0$ and $\sigma_{xy}^{IIIz}=\sigma_0^z$ if $\mu<0$. The latter relation is a well-known result for a spin hall conductivity~\cite{Sinova2004,Sinitsyn2004}. In a more general case, the results were obtained numerically and presented in Fig.~\ref{sxyzbulktopu}. As we can see from upper panel of this figure, the hexagonal warping has a little effect on the value of the isotropic spin conductivity, while the correction to the spin-orbit coupling $s$ enhances it. According to bottom panel of Fig.~\ref{sxyzbulktopu}, the anisotropic part of the spin conductivity, $\sigma_{xx}^{IIIx}$, decays monotonically with the increase of the chemical potential. It increases almost linearly with the increase of the hexagonal warping strength $\lambda$. The bulk spin conductivity becomes zero when the chemical potential crosses the bottom of the conduction band and occurs in the energy gap.

\begin{figure}[t!]
		\center
		\includegraphics [width=8.5cm, height=6cm]{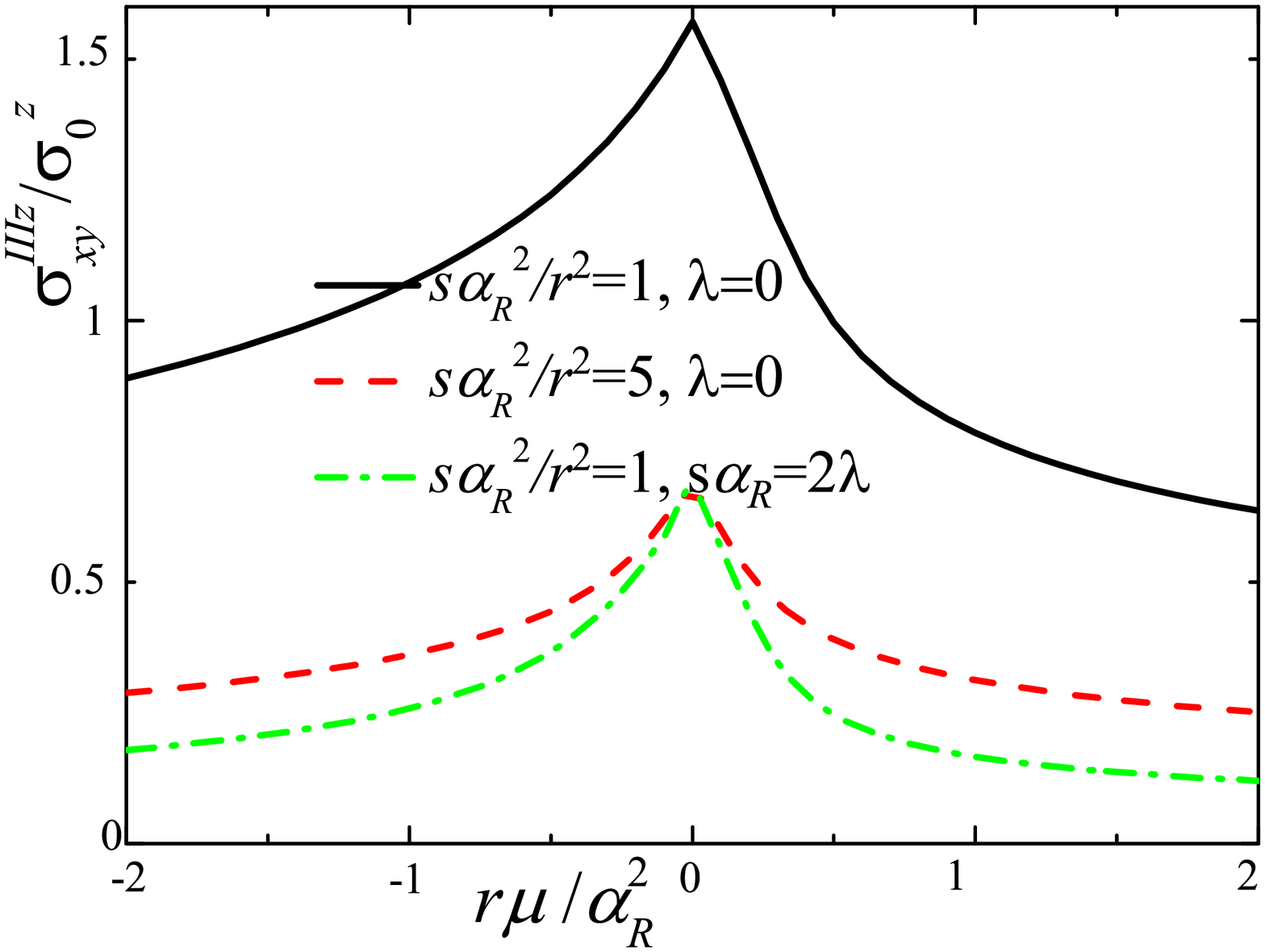}
			\includegraphics [width=8.5cm,height=6cm]{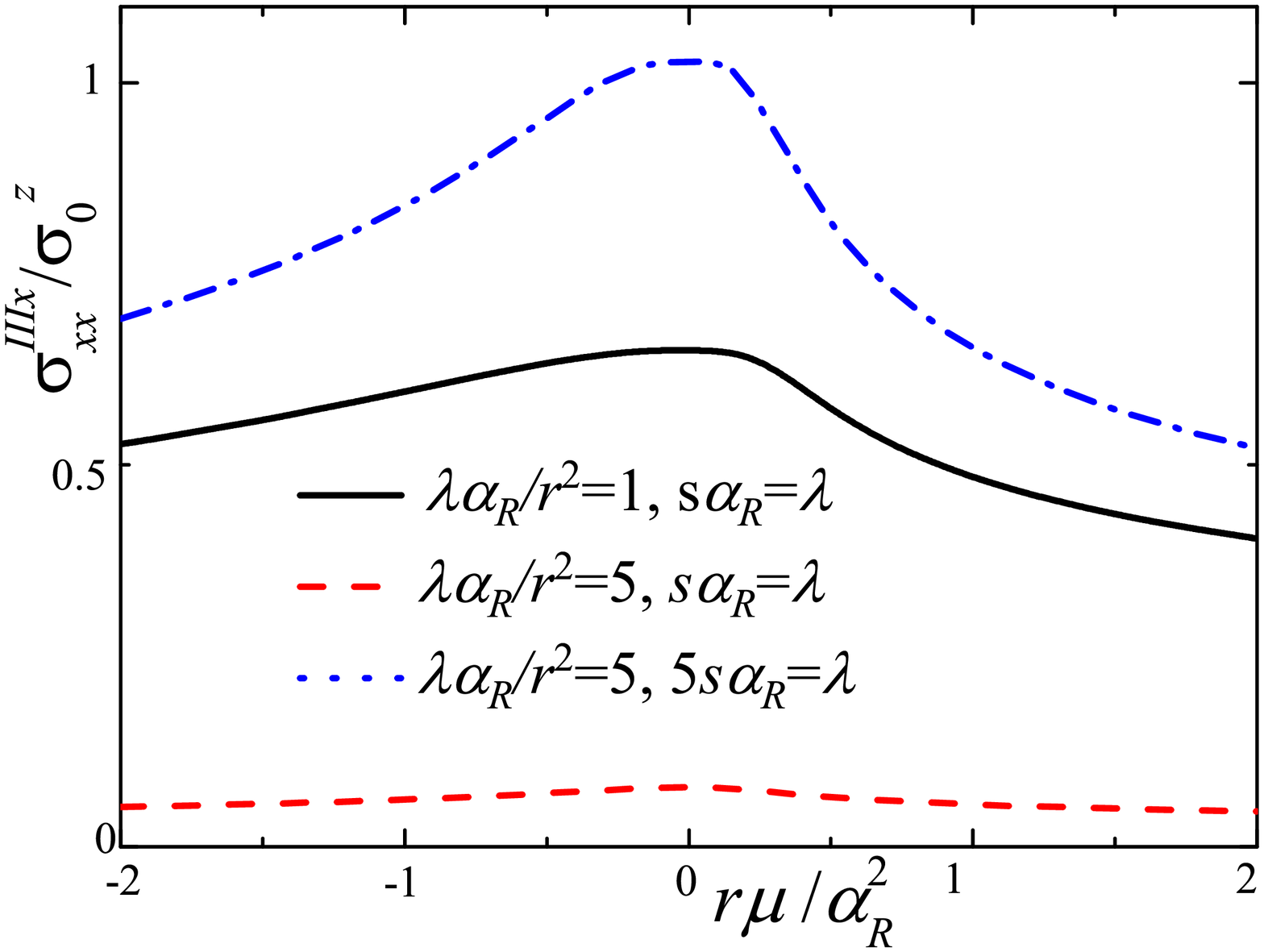}
		\caption{Isotropic spin conductivity $\sigma_{xy}^{IIIz}$ (top panel) and anisotropic spin conductivity $\sigma_{xx}^{IIIx}$ (bottom panel) for the surface states as a function of the chemical potential for $\Gamma=0$. Black line corresponds to $s\alpha_R^2/r^2=5$ and $\lambda=0$, red line to $s\alpha_R^2/r^2=1$ and $\lambda=0$, green line to $s\alpha_R^2/r^2=1$ and $2\lambda=s\alpha_R$.}\label{sxyzsurfacetopu}
	\end{figure}

The results for the contribution to the spin conductivity from the surface states are shown in Fig.~\ref{sxyzsurfacetopu}. We see that both $\sigma_{xy}^{IIIz}(\mu)$ and $\sigma_{xx}^{IIIz}(\mu)$ have maxima at the Dirac point $\mu=0$ and decreases with the increase of $|\mu|$ (in contrast to the contribution from the states at the Fermi level, Fig.~\ref{sxyzsurfaceu}). These functions are more or less particle-hole symmetric. From Fig.~\ref{sxyzsurfacetopu}, we see that the value of isotropic spin conductivity $\sigma_{xy}^{IIIz}$ decreases with the increase of the higher order momentum corrections $\lambda$ and $s$. The value of anisotropic spin conductivity $\sigma_{xx}^{IIIx}$ increases with the increase of hexagonal warping strength $\lambda$ and decreases with the increase of $s$.

\begin{figure}[t!]
\center
\includegraphics [width=8.5cm, height=6cm]{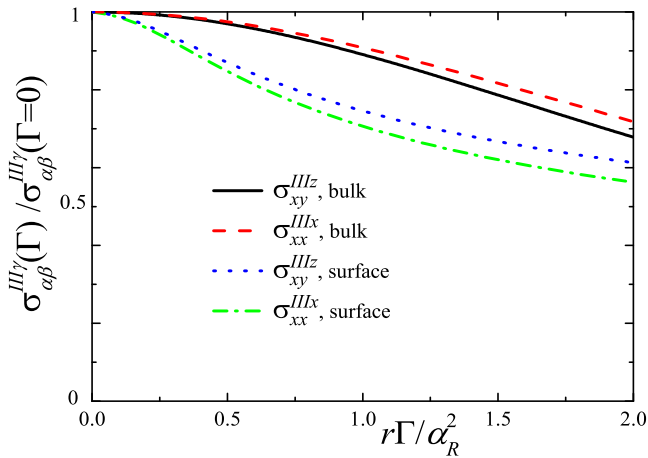}
\caption{Bulk and surface contributions to the spin conductivity from the filled states, $\sigma_{\alpha\beta}^{III\gamma}$, as a function of disorder parameter $\Gamma$ at $\mu=-\alpha_R^2/r$. Black line corresponds to the isotropic component due the bulk states, $\sigma_{xy}^{IIIz}$, for $s$ and $\lambda=0$, red line presents the anisotropic component due to the bulk states, $\sigma_{xx}^{IIIx}$, for $\lambda \alpha_R/r^2=0.1$ and $s=0$, blue line is the isotropic component due to the surface states, $\sigma_{xy}^{IIIz}$, for $s\alpha_R^2/r^2=5$ and $\lambda=0$, green line is the anisotropic component due to the surface states, $\sigma_{xx}^{IIIx}$, for $s\alpha_R^2/r^2=5$ and $\lambda =s\alpha_R$.}\label{topg}
	\end{figure}

The dependence of $\sigma_{\alpha \beta}^{III\gamma}$ on the disorder parameter $\Gamma$ is shown in Fig.~\ref{topg}. We obtain that the spin conductivity from the filled states is robust against disorder if $\gamma_b \ll 1$. If the disorder is stronger, $\gamma_b \sim 1$, both the bulk and surface conductivities are suppressed.

\section{Evaluation of characteristic parameters}\label{sec_evaluation}

In this section, we demonstrate that the values of the parameters used above for the calculation of the spin conductivity are reasonable.

We can extract information on the disorder strength from Ref.~\onlinecite{Chen2013}. The imaginary part of the self-energy for the surface states in Bi$_2$Te$_3$ can be estimated from the ARPES data presented in Ref.~\onlinecite{Chen2013} as a half-width of the quasiparticle peak: $\Gamma\approx 1$~meV and peak position corresponds to $\mu \approx 100$~meV. Thus, we get $\gamma_b \approx 10^{-2}$. This is an upper limit for the disorder strength, since, for example, electron-phonon and electron-electron interactions also contribute to the blurring of the quasiparticle peak. The alternative indirect estimate we obtain as follows. The STM data from Ref.~\onlinecite{Cheng2010} shows that for a clean surface of Bi$_2$Te$_3$ there exists one defect approximately per $\AA^2$. We suppose that a typical impurity potential is of the order of the chemical potential $\mu$ (which was about 200~meV). This assumption is true, e.g., for vacations. The Fermi velocity for the surface states was evaluated in Ref.~\onlinecite{Zhang2009} as $\alpha_R \approx 3$ eV$\cdot\AA^{-1}$. Then, we get $\gamma_b \approx 10^{-3}-10^{-2}$. The value of the Rashba spin-orbit coupling is comparable for the bulk and surface states~\cite{King2011}. Therefore, the value of $\alpha_R$ for the bulk states can be almost the same as for the surface states. Also, the electron mass for the bulk states~\cite{Piot2016} is close to that for the surface states~\cite{Nomura2014}. Thus it is reasonable to expect that $\gamma_b$ for the bulk states would of the same order as for the surface states.

In Table~\ref{table1} we put estimated values of the dimensionless parameters for Bi$_2$Se$_3$. These values were obtained by fitting the ARPES data presented in Refs.~\onlinecite{Nomura2014} and~\onlinecite{Rakyta2012}. Here subscripts $b$ and $s$ stands for the bulk and surface states, respectively. In general, the positions of the Dirac cone for the surface states is different from the position of zero $\mu$ for the bulk states (as it was defined in Fig.~\ref{spectra}). Thus, $\mu_s\neq\mu_b$. Naturally, we can extract reliable values of parameters $\alpha_R$, $r$, $\lambda$, and $s$ only for the surface states. We assume that the characteristics $\alpha_R$, $r$, and $\lambda$ are the same for the surface states and bulk states, while $s_b=0.1s_s$. We believe that such a choice does not affect the results within an order of magnitude.

We calculate the components of the spin conductivity for the set of parameters from Table~\ref{table1} and for the dimensionless disorder strength $\gamma_b = 10^{-3}$ estimated above. The results are presented in Table~\ref{table2}. We see that typically the isotropic, $\sigma_{xy}^z$, and anisotropic, $\sigma_{xx}^x$, components of the spin conductivity has the same order of magnitude. The contribution from the states at the Fermi level is comparable to the contribution from the filled states. The spin conductivity of the surface states is of the same order as the conductivity from the bulk states per layer.

		\begin{table}[]
			\centering
\caption{Dimensionless parameters extracted from the experimental data of Refs.~\onlinecite{Nomura2014} and~\onlinecite{Rakyta2012}.}
			\label{table1}
			\begin{tabular}{|l|l|l|l|l|l|l|l|l|}
				\hline
				& $\lambda \alpha_R/r^2$
				& $r\mu_{b}/\alpha_R^2$
				& $r\mu_{s}/\alpha_R^2$
				& $s_{b} \alpha_R^2/r^2$
				& $s_{s} \alpha_R^2/r^2$
				\\ \hline
				Bi$_2$Se$_3$ & 0.1  &0.1&  -0.5&0.07& 0.7  \\ \hline
			\end{tabular}
		\end{table}

	\begin{table}[]
		\centering
		\caption{Components of the spin conductivity calculated using the parameters from Table\ref{table1} and $\gamma_b = 10^{-3}$.}
		\label{table2}
		\begin{tabular}{|l|l|l|l|l|}
			\hline
			& $\sigma_{xy}^{Iz}/\sigma_0^z$
			& $\sigma_{xx}^{Ix}/\sigma_0^z$
			& $\sigma_{xy}^{IIIz}/\sigma_0^z$
			& $\sigma_{xx}^{IIIx}/\sigma_0^z$
			\\ \hline
			Bulk states per layer & 1.67 & 1.47 & 0.5 & 0.22 \\ \hline
			Surface states &  0.57 & 0.07  &  0.8 &  1.02  \\ \hline
		\end{tabular}
	\end{table}

\begin{figure}[t!]
		\center
		\includegraphics [width=8.5cm, height=6cm]{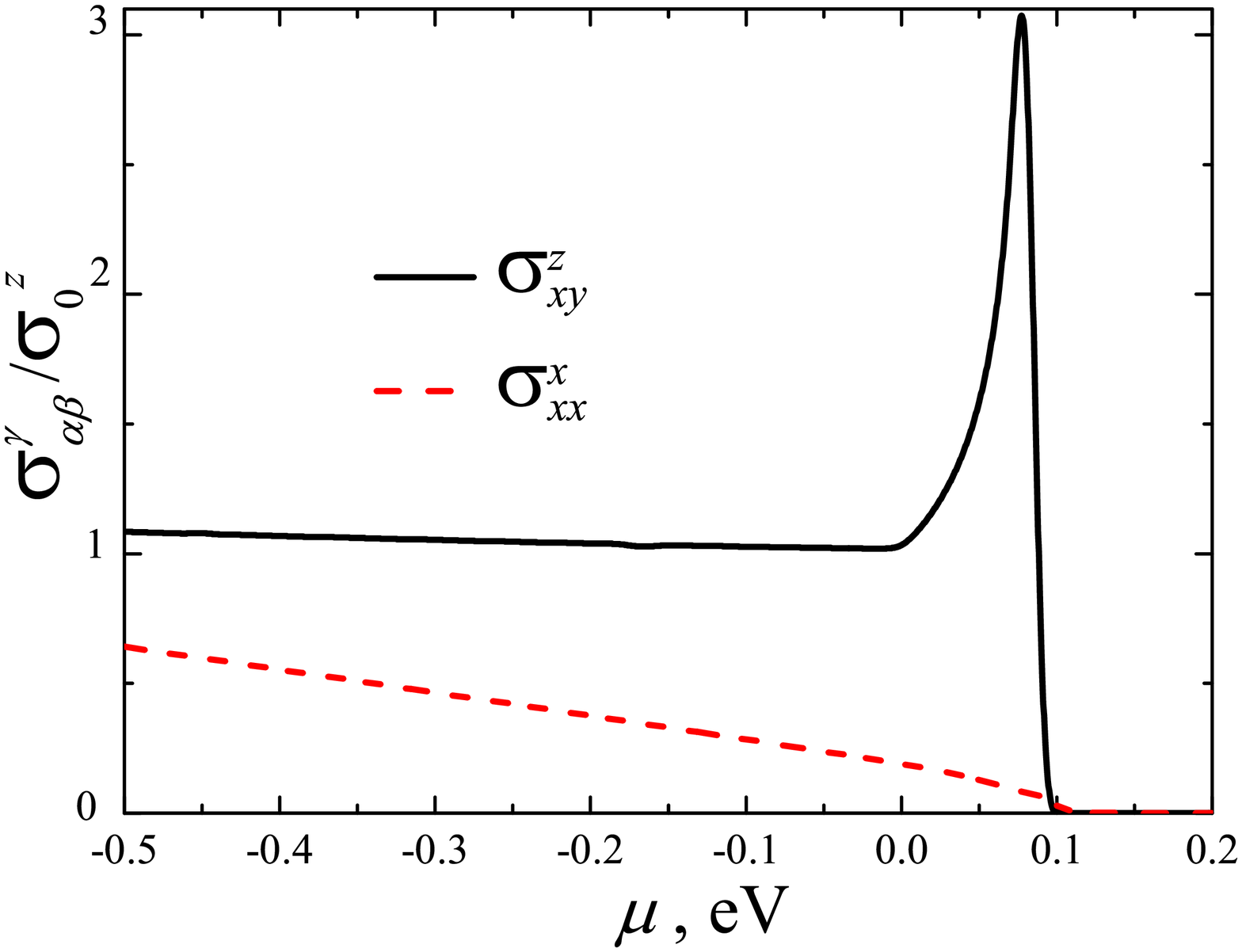}
				\includegraphics [width=8.5cm, height=6cm]{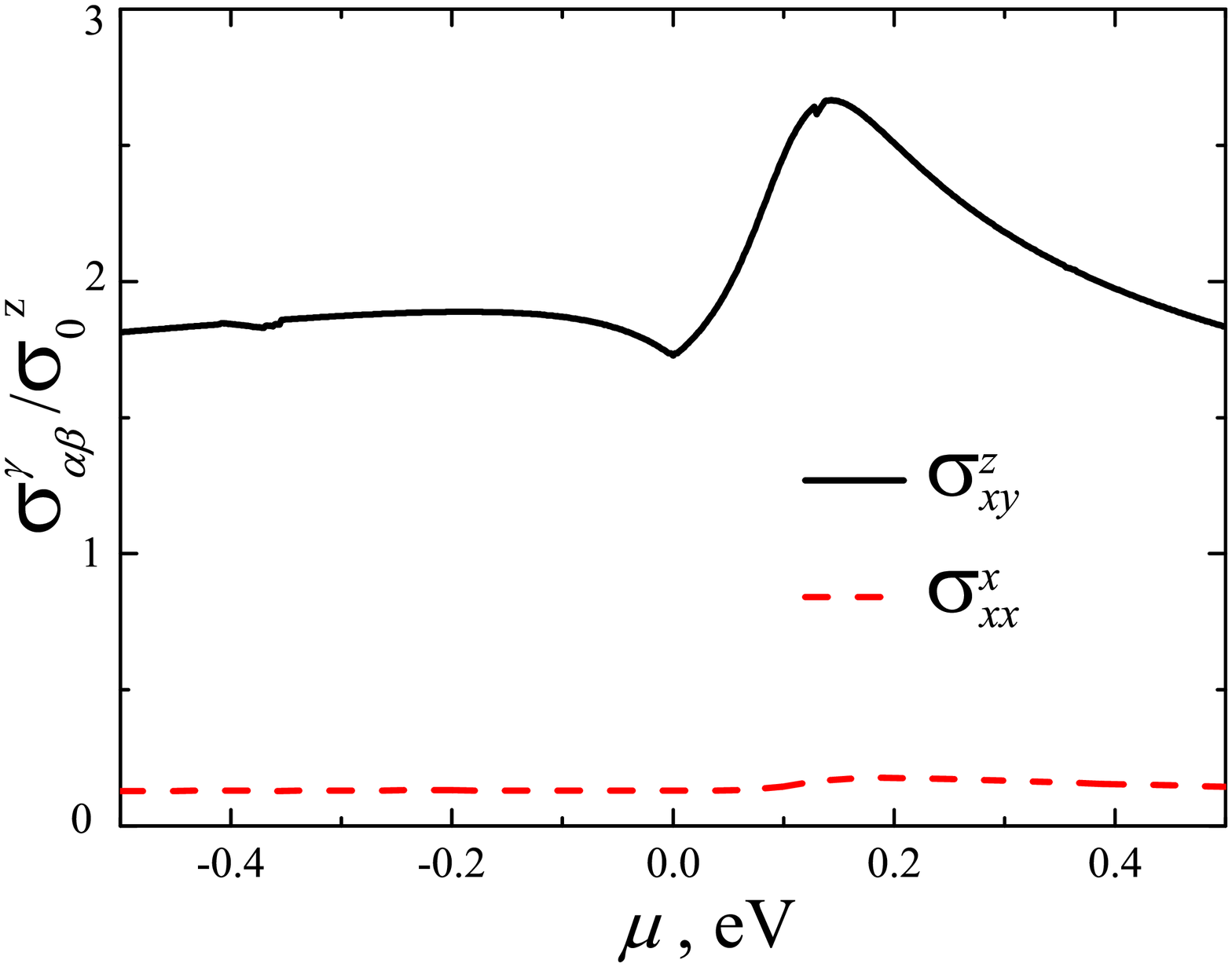}
		\caption{Total bulk spin conductivity per layer (top panel) and total surface spin conductivity (bottom panel) calculated for parameters from Table~\ref{table1} and $\gamma_b = 10^{-3}$. Black line shows isotropic spin conductivity $\sigma_{xy}^{z}$ and red line shows anisotropic component $\sigma_{xx}^{x}$.}\label{bi2te3bulk}
	\end{figure}

Using the estimated above values of parameters, we calculate the dependence of the total spin conductivity, $\sigma_{\alpha\beta}^{\gamma}=\sigma_{\alpha\beta}^{I\gamma}+\sigma_{\alpha\beta}^{III\gamma}$, on the chemical potential. The results for the bulk and surface  states are shown in Figs.~\ref{bi2te3bulk}. As we can see from the top panel in Fig.~\ref{bi2te3bulk}, the isotropic bulk spin conductivity slowly decreases from the plateau with an increase of $\mu$, then, has a peak, and finally drops to zero. It is rather high in the helical state. As for the anisotropic component, $\sigma_{xx}^{x}$, its value is considerably smaller than $\sigma_{xy}^{z}$. It monotonously decreases to zero with the growth of the chemical potential. The spin conductivity due to the surface states does not depend crucially on the chemical potential, Fig.~\ref{bi2te3bulk}. The anisotropic component of the surface spin conductivity is much smaller than the isotropic one.

	\section{Discussion}\label{sec_discuss}

The spin conductivity quanta can be expressed in the dimensional units as $\sigma_0^z \approx \left(\hbar /2e\right)\, 2 \cdot 10^{-5}\, \Omega^{-1}$. In the previous section we estimate that the spin conductivity per conducting layer (either bulk or surface) is of the order of $\sigma_0^z$. The distance between the layers in the TIs $l$ is of the order of 1~nm. Then, the specific (volume) spin conductivity can be estimated as $\sigma_{xy}^{z}l~\sim\sigma_{xx}^{x}l \sim \left(\hbar /2e\right) \, 2 \cdot 10^{4} \,\Omega^{-1}$m$^{-1}$. These values are close to that measured in Bi$_2$Se$_3$~\cite{Han2017}.

In the present study, however, we can not explain a colossal spin conductivity in BiSb~\cite{2017arXiv170907684H} and $($Bi$_{0.5}$Sb$_{0.5})_2$Te$_3$~\cite{Fan2014}, which is about two orders of magnitude higher than our estimation. This discrepancy may occur for the following reasons. First, it has been argued in Ref.~\onlinecite{Han2017} that different values of the spin conductivity can be a result of different fitting procedures of the experimental data. Second, in our consideration we do not take into account the effects of a magnetic field. The external magnetization (which is typically presented in the experiments~\cite{Fan2014}) could drastically enhance the spin conductivity.

In Refs.~\cite{Inoue2004,Raimondi2005} it has been shown that the vertex corrections in Rashba spin-orbit materials are small and, consequently, the (bulk in the TIs) spin conductivity from the states at the Fermi level is damped. However, in these papers the helical state, $0<\mu<\alpha_R^2/4r$, has not been considered. According to our analysis, in such phase, the vertex correction to the Rashba spin-orbit coupling is comparable to its bare value $\alpha_R$. So, a quite significant contribution to the spin current can be observed even from the bulk states. The vertex correction also increases the contribution from the surface states to the spin conductivity.

It has been speculated that the surface states can generate large spin currents observed in the experiment~\cite{Shiomi2014,Wang2015a,Wang2016}. According to our study, the surface states cannot produce very large spin current and the spin conductivity of the surface states typically has the same value as the spin conductivity of the bulk states per layer. So, our work confirms the experiments that show that spin conductivity mainly arises from the bulk states for a multilayer TI~\cite{Jamali2015,Han2017}  Also, it can explain the experiment, where the spin conductivity is small when the bulk of the TI sample is insulating, and the spin conductivity increases, when the bulk is conducting~\cite{Kondou2016}.

The bulk spin conductivity is robust against disorder if the spin-orbit coupling is large in comparison with a disorder, $\gamma_b \ll 1$. Otherwise, these contributions to the spin conductivity is suppressed, see Figs.~\ref{fermig} and~\ref{topg}. The surface spin conductivity is robust against disorder even if a disorder is not weak, $\gamma_b \sim 1$. The nature of robustness of the surface spin conductivity is similar to the robustness of the surface charge conductivity against disorder and arises due to suppression of the back-scattering. However, the study of the spin conductivity of the surface states in case of strong disorder deserves future studies.

The surface spin conductivity in a thin layer of TI with a cubic lattice have been studied in Ref.~\onlinecite{Peng2016} without taking into account the vertex corrections. The authors of the latter paper argue that the dependence of the surface spin conductivity on the disorder and chemical potential is weak. Our analysis confirms these results, see Figs.~\ref{fermig}, ~\ref{topg}, and~\ref{bi2te3bulk}.

The bulk spin conductivity can be tuned by a changing the chemical potential. Adjusting the chemical potential to the vicinity of the bottom of electron band, $\mu=\alpha_R^2/4r$, we can attain the largest spin currents, see Fig.~\ref{bi2te3bulk}. The tuning of the spin conductivity contribution from the filled states by changing the chemical potential has been demonstrated numerically in Ref.~\onlinecite{Sahin2015}. The dependence of the spin-hall angle on the chemical potential has been measured in the experiments~\cite{Fan2016}. However, the observed result can be explained not only by the present analysis but also by the particle-hole asymmetry of the charge current.

In the experiment, the components of the spin conductivity tensor are measure in the coordinate axes $\bar{x},\bar{y}$ related to the current. In Ref.~\onlinecite{Yang2016} the spin conductivity component $\sigma_{\bar{x}\bar{x}}^y$ was measured in BiSbTeSe$_2$. The authors observed different sign of this value in different samples. According to the conductivity tensor transformation presented in Section~\ref{sec_fermi}, Eq.~\eqref{new_axes}, the measured conductivity should be anisotropic and depends on the angle $\theta$ between the current and crystallographic $x$-axis as $\sigma_{\bar{x}\bar{x}}^y=-\sigma_{xx}^x\sin{3\theta}$. Thus, different sign of the measured spin conductivity may be due to the different orientation of the current leads with respect to the crystallographic axes in different samples.

\section*{Acknowledgements}

We acknowledge support from the Russian Scientific Foundation, Grant No 17-12-01544. RSA acknowledge the partial support by the Basis Foundation and ICFPM (MMK) of Education and Science of the Russian Federation, Grant No. 14Y26.31.0007.

	\bibliographystyle{apsrevlong_no_issn_url}
	\bibliography{bib_hw}

\begin{thebibliography}{49}
\expandafter\ifx\csname natexlab\endcsname\relax\def\natexlab#1{#1}\fi
\expandafter\ifx\csname bibnamefont\endcsname\relax
  \def\bibnamefont#1{#1}\fi
\expandafter\ifx\csname bibfnamefont\endcsname\relax
  \def\bibfnamefont#1{#1}\fi
\expandafter\ifx\csname citenamefont\endcsname\relax
  \def\citenamefont#1{#1}\fi

\bibitem[{\citenamefont{Hasan and Kane}(2010)}]{RevModPhys.82.3045}
\bibinfo{author}{\bibfnamefont{M.~Z.} \bibnamefont{Hasan}} \bibnamefont{and}
  \bibinfo{author}{\bibfnamefont{C.~L.} \bibnamefont{Kane}},
  {``}\bibinfo{title}{Colloquium: Topological insulators},{''}
  \bibinfo{journal}{Rev. Mod. Phys.} \textbf{\bibinfo{volume}{82}},
  \bibinfo{pages}{3045} (\bibinfo{year}{2010}).

\bibitem[{\citenamefont{Fu}(2009)}]{Fu2009}
\bibinfo{author}{\bibfnamefont{L.}~\bibnamefont{Fu}},
  {``}\bibinfo{title}{Hexagonal Warping Effects in the Surface States of the
  Topological Insulator ${\mathrm{Bi}}_{2}{\mathrm{Te}}_{3}$},{''}
  \bibinfo{journal}{Phys. Rev. Lett.} \textbf{\bibinfo{volume}{103}},
  \bibinfo{pages}{266801} (\bibinfo{year}{2009}).

\bibitem[{\citenamefont{Kuroda et~al.}(2010)\citenamefont{Kuroda, Arita,
  Miyamoto, Ye, Jiang, Kimura, Krasovskii, Chulkov, Iwasawa, Okuda
  et~al.}}]{Kuroda2010}
\bibinfo{author}{\bibfnamefont{K.}~\bibnamefont{Kuroda}},
  \bibinfo{author}{\bibfnamefont{M.}~\bibnamefont{Arita}},
  \bibinfo{author}{\bibfnamefont{K.}~\bibnamefont{Miyamoto}},
  \bibinfo{author}{\bibfnamefont{M.}~\bibnamefont{Ye}},
  \bibinfo{author}{\bibfnamefont{J.}~\bibnamefont{Jiang}},
  \bibinfo{author}{\bibfnamefont{A.}~\bibnamefont{Kimura}},
  \bibinfo{author}{\bibfnamefont{E.~E.} \bibnamefont{Krasovskii}},
  \bibinfo{author}{\bibfnamefont{E.~V.} \bibnamefont{Chulkov}},
  \bibinfo{author}{\bibfnamefont{H.}~\bibnamefont{Iwasawa}},
  \bibinfo{author}{\bibfnamefont{T.}~\bibnamefont{Okuda}},
  \bibnamefont{et~al.}, {``}\bibinfo{title}{Hexagonally Deformed Fermi Surface
  of the 3D Topological Insulator ${\mathrm{Bi}}_{2}{\mathrm{Se}}_{3}$},{''}
  \bibinfo{journal}{Phys. Rev. Lett.} \textbf{\bibinfo{volume}{105}},
  \bibinfo{pages}{076802} (\bibinfo{year}{2010}).

\bibitem[{\citenamefont{Wang and Yu}(2011)}]{Wang2011}
\bibinfo{author}{\bibfnamefont{C.~M.} \bibnamefont{Wang}} \bibnamefont{and}
  \bibinfo{author}{\bibfnamefont{F.~J.} \bibnamefont{Yu}},
  {``}\bibinfo{title}{Effects of hexagonal warping on surface transport in
  topological insulators},{''} \bibinfo{journal}{Phys. Rev. B}
  \textbf{\bibinfo{volume}{84}}, \bibinfo{pages}{155440}
  (\bibinfo{year}{2011}).

\bibitem[{\citenamefont{Pal et~al.}(2012)\citenamefont{Pal, Yudson, and
  Maslov}}]{Pal2012}
\bibinfo{author}{\bibfnamefont{H.~K.} \bibnamefont{Pal}},
  \bibinfo{author}{\bibfnamefont{V.~I.} \bibnamefont{Yudson}},
  \bibnamefont{and} \bibinfo{author}{\bibfnamefont{D.~L.}
  \bibnamefont{Maslov}}, {``}\bibinfo{title}{Effect of electron-electron
  interaction on surface transport in the Bi${}_{2}$Te${}_{3}$ family of
  three-dimensional topological insulators},{''} \bibinfo{journal}{Phys. Rev.
  B} \textbf{\bibinfo{volume}{85}}, \bibinfo{pages}{085439}
  (\bibinfo{year}{2012}).

\bibitem[{\citenamefont{Siu et~al.}(2014)\citenamefont{Siu, Jalil, and
  Tan}}]{Siu2014}
\bibinfo{author}{\bibfnamefont{Z.~B.} \bibnamefont{Siu}},
  \bibinfo{author}{\bibfnamefont{M.~B.~A.} \bibnamefont{Jalil}},
  \bibnamefont{and} \bibinfo{author}{\bibfnamefont{S.~G.} \bibnamefont{Tan}},
  {``}\bibinfo{title}{Topological state transport in topological insulators
  under the influence of hexagonal warping and exchange coupling to in-plane
  magnetizations},{''} \bibinfo{journal}{Scientific Reports}
  \textbf{\bibinfo{volume}{4}}, \bibinfo{pages}{5062} (\bibinfo{year}{2014}).

\bibitem[{\citenamefont{Repin and Burmistrov}(2015)}]{Repin2015}
\bibinfo{author}{\bibfnamefont{E.~V.} \bibnamefont{Repin}} \bibnamefont{and}
  \bibinfo{author}{\bibfnamefont{I.~S.} \bibnamefont{Burmistrov}},
  {``}\bibinfo{title}{Surface states in a 3D topological insulator: The role of
  hexagonal warping and curvature},{''} \bibinfo{journal}{Journal of
  Experimental and Theoretical Physics} \textbf{\bibinfo{volume}{121}},
  \bibinfo{pages}{509} (\bibinfo{year}{2015}).

\bibitem[{\citenamefont{Akzyanov and Rakhmanov}(2018)}]{PhysRevB.97.075421}
\bibinfo{author}{\bibfnamefont{R.~S.} \bibnamefont{Akzyanov}} \bibnamefont{and}
  \bibinfo{author}{\bibfnamefont{A.~L.} \bibnamefont{Rakhmanov}},
  {``}\bibinfo{title}{Surface charge conductivity of a topological insulator in
  a magnetic field: The effect of hexagonal warping},{''}
  \bibinfo{journal}{Phys. Rev. B} \textbf{\bibinfo{volume}{97}},
  \bibinfo{pages}{075421} (\bibinfo{year}{2018}).

\bibitem[{\citenamefont{Sinova et~al.}(2015)\citenamefont{Sinova, Valenzuela,
  Wunderlich, Back, and Jungwirth}}]{Sinova2015}
\bibinfo{author}{\bibfnamefont{J.}~\bibnamefont{Sinova}},
  \bibinfo{author}{\bibfnamefont{S.~O.} \bibnamefont{Valenzuela}},
  \bibinfo{author}{\bibfnamefont{J.}~\bibnamefont{Wunderlich}},
  \bibinfo{author}{\bibfnamefont{C.}~\bibnamefont{Back}}, \bibnamefont{and}
  \bibinfo{author}{\bibfnamefont{T.}~\bibnamefont{Jungwirth}},
  {``}\bibinfo{title}{Spin Hall effects},{''} \bibinfo{journal}{Rev. Mod.
  Phys.} \textbf{\bibinfo{volume}{87}}, \bibinfo{pages}{1213}
  (\bibinfo{year}{2015}).

\bibitem[{\citenamefont{Murakami et~al.}(2003)\citenamefont{Murakami, Nagaosa,
  and Zhang}}]{Murakami2003}
\bibinfo{author}{\bibfnamefont{S.}~\bibnamefont{Murakami}},
  \bibinfo{author}{\bibfnamefont{N.}~\bibnamefont{Nagaosa}}, \bibnamefont{and}
  \bibinfo{author}{\bibfnamefont{S.-C.} \bibnamefont{Zhang}},
  {``}\bibinfo{title}{Dissipationless Quantum Spin Current at Room
  Temperature},{''} \bibinfo{journal}{Science} \textbf{\bibinfo{volume}{301}},
  \bibinfo{pages}{1348} (\bibinfo{year}{2003}).

\bibitem[{\citenamefont{Sinova et~al.}(2004)\citenamefont{Sinova, Culcer, Niu,
  Sinitsyn, Jungwirth, and MacDonald}}]{Sinova2004}
\bibinfo{author}{\bibfnamefont{J.}~\bibnamefont{Sinova}},
  \bibinfo{author}{\bibfnamefont{D.}~\bibnamefont{Culcer}},
  \bibinfo{author}{\bibfnamefont{Q.}~\bibnamefont{Niu}},
  \bibinfo{author}{\bibfnamefont{N.~A.} \bibnamefont{Sinitsyn}},
  \bibinfo{author}{\bibfnamefont{T.}~\bibnamefont{Jungwirth}},
  \bibnamefont{and} \bibinfo{author}{\bibfnamefont{A.~H.}
  \bibnamefont{MacDonald}}, {``}\bibinfo{title}{Universal Intrinsic Spin Hall
  Effect},{''} \bibinfo{journal}{Phys. Rev. Lett.}
  \textbf{\bibinfo{volume}{92}}, \bibinfo{pages}{126603}
  (\bibinfo{year}{2004}).

\bibitem[{\citenamefont{Inoue et~al.}(2004)\citenamefont{Inoue, Bauer, and
  Molenkamp}}]{Inoue2004}
\bibinfo{author}{\bibfnamefont{J.-i.} \bibnamefont{Inoue}},
  \bibinfo{author}{\bibfnamefont{G.~E.~W.} \bibnamefont{Bauer}},
  \bibnamefont{and} \bibinfo{author}{\bibfnamefont{L.~W.}
  \bibnamefont{Molenkamp}}, {``}\bibinfo{title}{Suppression of the persistent
  spin Hall current by defect scattering},{''} \bibinfo{journal}{Phys. Rev. B}
  \textbf{\bibinfo{volume}{70}}, \bibinfo{pages}{041303}
  (\bibinfo{year}{2004}).

\bibitem[{\citenamefont{Raimondi and Schwab}(2005)}]{Raimondi2005}
\bibinfo{author}{\bibfnamefont{R.}~\bibnamefont{Raimondi}} \bibnamefont{and}
  \bibinfo{author}{\bibfnamefont{P.}~\bibnamefont{Schwab}},
  {``}\bibinfo{title}{Spin-Hall effect in a disordered two-dimensional electron
  system},{''} \bibinfo{journal}{Phys. Rev. B} \textbf{\bibinfo{volume}{71}},
  \bibinfo{pages}{033311} (\bibinfo{year}{2005}).

\bibitem[{\citenamefont{Ralph and Stiles}(2008)}]{Ralph2008}
\bibinfo{author}{\bibfnamefont{D.}~\bibnamefont{Ralph}} \bibnamefont{and}
  \bibinfo{author}{\bibfnamefont{M.}~\bibnamefont{Stiles}},
  {``}\bibinfo{title}{Spin transfer torques},{''} \bibinfo{journal}{Journal of
  Magnetism and Magnetic Materials} \textbf{\bibinfo{volume}{320}},
  \bibinfo{pages}{1190} (\bibinfo{year}{2008}).

\bibitem[{\citenamefont{Chen et~al.}(2015)\citenamefont{Chen, Sigrist, Sinova,
  and Manske}}]{Chen2015}
\bibinfo{author}{\bibfnamefont{W.}~\bibnamefont{Chen}},
  \bibinfo{author}{\bibfnamefont{M.}~\bibnamefont{Sigrist}},
  \bibinfo{author}{\bibfnamefont{J.}~\bibnamefont{Sinova}}, \bibnamefont{and}
  \bibinfo{author}{\bibfnamefont{D.}~\bibnamefont{Manske}},
  {``}\bibinfo{title}{Minimal Model of Spin-Transfer Torque and Spin Pumping
  Caused by the Spin Hall Effect},{''} \bibinfo{journal}{Phys. Rev. Lett.}
  \textbf{\bibinfo{volume}{115}}, \bibinfo{pages}{217203}
  (\bibinfo{year}{2015}).

\bibitem[{\citenamefont{Wang et~al.}(2013)\citenamefont{Wang, Alzate, and
  Amiri}}]{Wang2013}
\bibinfo{author}{\bibfnamefont{K.~L.} \bibnamefont{Wang}},
  \bibinfo{author}{\bibfnamefont{J.~G.} \bibnamefont{Alzate}},
  \bibnamefont{and} \bibinfo{author}{\bibfnamefont{P.~K.} \bibnamefont{Amiri}},
  {``}\bibinfo{title}{Low-power non-volatile spintronic memory: STT-RAM and
  beyond},{''} \bibinfo{journal}{Journal of Physics D: Applied Physics}
  \textbf{\bibinfo{volume}{46}}, \bibinfo{pages}{074003}
  (\bibinfo{year}{2013}).

\bibitem[{\citenamefont{Mellnik et~al.}(2014)\citenamefont{Mellnik, Lee,
  Richardella, Grab, Mintun, Fischer, Vaezi, Manchon, Kim, Samarth
  et~al.}}]{Mellnik2014}
\bibinfo{author}{\bibfnamefont{A.~R.} \bibnamefont{Mellnik}},
  \bibinfo{author}{\bibfnamefont{J.~S.} \bibnamefont{Lee}},
  \bibinfo{author}{\bibfnamefont{A.}~\bibnamefont{Richardella}},
  \bibinfo{author}{\bibfnamefont{J.~L.} \bibnamefont{Grab}},
  \bibinfo{author}{\bibfnamefont{P.~J.} \bibnamefont{Mintun}},
  \bibinfo{author}{\bibfnamefont{M.~H.} \bibnamefont{Fischer}},
  \bibinfo{author}{\bibfnamefont{A.}~\bibnamefont{Vaezi}},
  \bibinfo{author}{\bibfnamefont{A.}~\bibnamefont{Manchon}},
  \bibinfo{author}{\bibfnamefont{E.-A.} \bibnamefont{Kim}},
  \bibinfo{author}{\bibfnamefont{N.}~\bibnamefont{Samarth}},
  \bibnamefont{et~al.}, {``}\bibinfo{title}{Spin-transfer torque generated by a
  topological insulator},{''} \bibinfo{journal}{Nature}
  \textbf{\bibinfo{volume}{511}}, \bibinfo{pages}{449} (\bibinfo{year}{2014}).

\bibitem[{\citenamefont{Fan et~al.}(2014)\citenamefont{Fan, Upadhyaya, Kou,
  Lang, Takei, Wang, Tang, He, Chang, Montazeri et~al.}}]{Fan2014}
\bibinfo{author}{\bibfnamefont{Y.}~\bibnamefont{Fan}},
  \bibinfo{author}{\bibfnamefont{P.}~\bibnamefont{Upadhyaya}},
  \bibinfo{author}{\bibfnamefont{X.}~\bibnamefont{Kou}},
  \bibinfo{author}{\bibfnamefont{M.}~\bibnamefont{Lang}},
  \bibinfo{author}{\bibfnamefont{S.}~\bibnamefont{Takei}},
  \bibinfo{author}{\bibfnamefont{Z.}~\bibnamefont{Wang}},
  \bibinfo{author}{\bibfnamefont{J.}~\bibnamefont{Tang}},
  \bibinfo{author}{\bibfnamefont{L.}~\bibnamefont{He}},
  \bibinfo{author}{\bibfnamefont{L.-T.} \bibnamefont{Chang}},
  \bibinfo{author}{\bibfnamefont{M.}~\bibnamefont{Montazeri}},
  \bibnamefont{et~al.}, {``}\bibinfo{title}{Magnetization switching through
  giant spin–orbit torque in a magnetically doped topological insulator
  heterostructure},{''} \bibinfo{journal}{Nat Mater}
  \textbf{\bibinfo{volume}{13}}, \bibinfo{pages}{699} (\bibinfo{year}{2014}).

\bibitem[{\citenamefont{Fan et~al.}(2016)\citenamefont{Fan, Kou, Upadhyaya,
  Shao, Pan, Lang, Che, Tang, Montazeri, Murata et~al.}}]{Fan2016}
\bibinfo{author}{\bibfnamefont{Y.}~\bibnamefont{Fan}},
  \bibinfo{author}{\bibfnamefont{X.}~\bibnamefont{Kou}},
  \bibinfo{author}{\bibfnamefont{P.}~\bibnamefont{Upadhyaya}},
  \bibinfo{author}{\bibfnamefont{Q.}~\bibnamefont{Shao}},
  \bibinfo{author}{\bibfnamefont{L.}~\bibnamefont{Pan}},
  \bibinfo{author}{\bibfnamefont{M.}~\bibnamefont{Lang}},
  \bibinfo{author}{\bibfnamefont{X.}~\bibnamefont{Che}},
  \bibinfo{author}{\bibfnamefont{J.}~\bibnamefont{Tang}},
  \bibinfo{author}{\bibfnamefont{M.}~\bibnamefont{Montazeri}},
  \bibinfo{author}{\bibfnamefont{K.}~\bibnamefont{Murata}},
  \bibnamefont{et~al.}, {``}\bibinfo{title}{Electric-field control of
  spin–orbit torque in a magnetically doped topological insulator},{''}
  \bibinfo{journal}{Nat Nano} \textbf{\bibinfo{volume}{11}},
  \bibinfo{pages}{352} (\bibinfo{year}{2016}).

\bibitem[{\citenamefont{Han et~al.}(2017)\citenamefont{Han, Richardella,
  Siddiqui, Finley, Samarth, and Liu}}]{Han2017}
\bibinfo{author}{\bibfnamefont{J.}~\bibnamefont{Han}},
  \bibinfo{author}{\bibfnamefont{A.}~\bibnamefont{Richardella}},
  \bibinfo{author}{\bibfnamefont{S.~A.} \bibnamefont{Siddiqui}},
  \bibinfo{author}{\bibfnamefont{J.}~\bibnamefont{Finley}},
  \bibinfo{author}{\bibfnamefont{N.}~\bibnamefont{Samarth}}, \bibnamefont{and}
  \bibinfo{author}{\bibfnamefont{L.}~\bibnamefont{Liu}},
  {``}\bibinfo{title}{Room-Temperature Spin-Orbit Torque Switching Induced by a
  Topological Insulator},{''} \bibinfo{journal}{Phys. Rev. Lett.}
  \textbf{\bibinfo{volume}{119}}, \bibinfo{pages}{077702}
  (\bibinfo{year}{2017}).

\bibitem[{\citenamefont{Yang et~al.}(2016)\citenamefont{Yang, Ghatak, Taskin,
  Segawa, Ando, Shiraishi, Kanai, Matsumoto, Rosch, and Ando}}]{Yang2016}
\bibinfo{author}{\bibfnamefont{F.}~\bibnamefont{Yang}},
  \bibinfo{author}{\bibfnamefont{S.}~\bibnamefont{Ghatak}},
  \bibinfo{author}{\bibfnamefont{A.~A.} \bibnamefont{Taskin}},
  \bibinfo{author}{\bibfnamefont{K.}~\bibnamefont{Segawa}},
  \bibinfo{author}{\bibfnamefont{Y.}~\bibnamefont{Ando}},
  \bibinfo{author}{\bibfnamefont{M.}~\bibnamefont{Shiraishi}},
  \bibinfo{author}{\bibfnamefont{Y.}~\bibnamefont{Kanai}},
  \bibinfo{author}{\bibfnamefont{K.}~\bibnamefont{Matsumoto}},
  \bibinfo{author}{\bibfnamefont{A.}~\bibnamefont{Rosch}}, \bibnamefont{and}
  \bibinfo{author}{\bibfnamefont{Y.}~\bibnamefont{Ando}},
  {``}\bibinfo{title}{Switching of charge-current-induced spin polarization in
  the topological insulator ${\mathrm{BiSbTeSe}}_{2}$},{''}
  \bibinfo{journal}{Phys. Rev. B} \textbf{\bibinfo{volume}{94}},
  \bibinfo{pages}{075304} (\bibinfo{year}{2016}).

\bibitem[{\citenamefont{Kondou et~al.}(2016)\citenamefont{Kondou, Yoshimi,
  Tsukazaki, Fukuma, Matsuno, Takahashi, Kawasaki, Tokura, and
  Otani}}]{Kondou2016}
\bibinfo{author}{\bibfnamefont{K.}~\bibnamefont{Kondou}},
  \bibinfo{author}{\bibfnamefont{R.}~\bibnamefont{Yoshimi}},
  \bibinfo{author}{\bibfnamefont{A.}~\bibnamefont{Tsukazaki}},
  \bibinfo{author}{\bibfnamefont{Y.}~\bibnamefont{Fukuma}},
  \bibinfo{author}{\bibfnamefont{J.}~\bibnamefont{Matsuno}},
  \bibinfo{author}{\bibfnamefont{K.~S.} \bibnamefont{Takahashi}},
  \bibinfo{author}{\bibfnamefont{M.}~\bibnamefont{Kawasaki}},
  \bibinfo{author}{\bibfnamefont{Y.}~\bibnamefont{Tokura}}, \bibnamefont{and}
  \bibinfo{author}{\bibfnamefont{Y.}~\bibnamefont{Otani}},
  {``}\bibinfo{title}{Fermi-level-dependent charge-to-spin current conversion
  by Dirac surface states of topological insulators},{''} \bibinfo{journal}{Nat
  Phys} \textbf{\bibinfo{volume}{12}}, \bibinfo{pages}{1027}
  (\bibinfo{year}{2016}).

\bibitem[{\citenamefont{Shiomi et~al.}(2014)\citenamefont{Shiomi, Nomura,
  Kajiwara, Eto, Novak, Segawa, Ando, and Saitoh}}]{Shiomi2014}
\bibinfo{author}{\bibfnamefont{Y.}~\bibnamefont{Shiomi}},
  \bibinfo{author}{\bibfnamefont{K.}~\bibnamefont{Nomura}},
  \bibinfo{author}{\bibfnamefont{Y.}~\bibnamefont{Kajiwara}},
  \bibinfo{author}{\bibfnamefont{K.}~\bibnamefont{Eto}},
  \bibinfo{author}{\bibfnamefont{M.}~\bibnamefont{Novak}},
  \bibinfo{author}{\bibfnamefont{K.}~\bibnamefont{Segawa}},
  \bibinfo{author}{\bibfnamefont{Y.}~\bibnamefont{Ando}}, \bibnamefont{and}
  \bibinfo{author}{\bibfnamefont{E.}~\bibnamefont{Saitoh}},
  {``}\bibinfo{title}{Spin-Electricity Conversion Induced by Spin Injection
  into Topological Insulators},{''} \bibinfo{journal}{Phys. Rev. Lett.}
  \textbf{\bibinfo{volume}{113}}, \bibinfo{pages}{196601}
  (\bibinfo{year}{2014}).

\bibitem[{\citenamefont{Wang et~al.}(2015)\citenamefont{Wang, Deorani,
  Banerjee, Koirala, Brahlek, Oh, and Yang}}]{Wang2015a}
\bibinfo{author}{\bibfnamefont{Y.}~\bibnamefont{Wang}},
  \bibinfo{author}{\bibfnamefont{P.}~\bibnamefont{Deorani}},
  \bibinfo{author}{\bibfnamefont{K.}~\bibnamefont{Banerjee}},
  \bibinfo{author}{\bibfnamefont{N.}~\bibnamefont{Koirala}},
  \bibinfo{author}{\bibfnamefont{M.}~\bibnamefont{Brahlek}},
  \bibinfo{author}{\bibfnamefont{S.}~\bibnamefont{Oh}}, \bibnamefont{and}
  \bibinfo{author}{\bibfnamefont{H.}~\bibnamefont{Yang}},
  {``}\bibinfo{title}{Topological Surface States Originated Spin-Orbit Torques
  in ${\mathrm{Bi}}_{2}{\mathrm{Se}}_{3}$},{''} \bibinfo{journal}{Phys. Rev.
  Lett.} \textbf{\bibinfo{volume}{114}}, \bibinfo{pages}{257202}
  (\bibinfo{year}{2015}).

\bibitem[{\citenamefont{Wang et~al.}(2016)\citenamefont{Wang, Kally, Lee, Liu,
  Chang, Hickey, Mkhoyan, Wu, Richardella, and Samarth}}]{Wang2016}
\bibinfo{author}{\bibfnamefont{H.}~\bibnamefont{Wang}},
  \bibinfo{author}{\bibfnamefont{J.}~\bibnamefont{Kally}},
  \bibinfo{author}{\bibfnamefont{J.~S.} \bibnamefont{Lee}},
  \bibinfo{author}{\bibfnamefont{T.}~\bibnamefont{Liu}},
  \bibinfo{author}{\bibfnamefont{H.}~\bibnamefont{Chang}},
  \bibinfo{author}{\bibfnamefont{D.~R.} \bibnamefont{Hickey}},
  \bibinfo{author}{\bibfnamefont{K.~A.} \bibnamefont{Mkhoyan}},
  \bibinfo{author}{\bibfnamefont{M.}~\bibnamefont{Wu}},
  \bibinfo{author}{\bibfnamefont{A.}~\bibnamefont{Richardella}},
  \bibnamefont{and} \bibinfo{author}{\bibfnamefont{N.}~\bibnamefont{Samarth}},
  {``}\bibinfo{title}{Surface-State-Dominated Spin-Charge Current Conversion in
  Topological-Insulator--Ferromagnetic-Insulator Heterostructures},{''}
  \bibinfo{journal}{Phys. Rev. Lett.} \textbf{\bibinfo{volume}{117}},
  \bibinfo{pages}{076601} (\bibinfo{year}{2016}).

\bibitem[{\citenamefont{Jamali et~al.}(2015)\citenamefont{Jamali, Lee, Jeong,
  Mahfouzi, Lv, Zhao, Nikolić, Mkhoyan, Samarth, and Wang}}]{Jamali2015}
\bibinfo{author}{\bibfnamefont{M.}~\bibnamefont{Jamali}},
  \bibinfo{author}{\bibfnamefont{J.~S.} \bibnamefont{Lee}},
  \bibinfo{author}{\bibfnamefont{J.~S.} \bibnamefont{Jeong}},
  \bibinfo{author}{\bibfnamefont{F.}~\bibnamefont{Mahfouzi}},
  \bibinfo{author}{\bibfnamefont{Y.}~\bibnamefont{Lv}},
  \bibinfo{author}{\bibfnamefont{Z.}~\bibnamefont{Zhao}},
  \bibinfo{author}{\bibfnamefont{B.~K.} \bibnamefont{Nikolić}},
  \bibinfo{author}{\bibfnamefont{K.~A.} \bibnamefont{Mkhoyan}},
  \bibinfo{author}{\bibfnamefont{N.}~\bibnamefont{Samarth}}, \bibnamefont{and}
  \bibinfo{author}{\bibfnamefont{J.-P.} \bibnamefont{Wang}},
  {``}\bibinfo{title}{Giant Spin Pumping and Inverse Spin Hall Effect in the
  Presence of Surface and Bulk Spin−Orbit Coupling of Topological Insulator
  Bi2Se3},{''} \bibinfo{journal}{Nano Lett.} \textbf{\bibinfo{volume}{15}},
  \bibinfo{pages}{7126} (\bibinfo{year}{2015}).

\bibitem[{\citenamefont{Yang and Chang}(2006)}]{Yang2006}
\bibinfo{author}{\bibfnamefont{M.-F.} \bibnamefont{Yang}} \bibnamefont{and}
  \bibinfo{author}{\bibfnamefont{M.-C.} \bibnamefont{Chang}},
  {``}\bibinfo{title}{St\ifmmode \check{r}\else \v{r}\fi{}eda-like formula in
  the spin Hall effect},{''} \bibinfo{journal}{Phys. Rev. B}
  \textbf{\bibinfo{volume}{73}}, \bibinfo{pages}{073304}
  (\bibinfo{year}{2006}).

\bibitem[{\citenamefont{Kodderitzsch et~al.}(2015)\citenamefont{Kodderitzsch,
  Chadova, and Ebert}}]{Kodderitzsch2015}
\bibinfo{author}{\bibfnamefont{D.}~\bibnamefont{Kodderitzsch}},
  \bibinfo{author}{\bibfnamefont{K.}~\bibnamefont{Chadova}}, \bibnamefont{and}
  \bibinfo{author}{\bibfnamefont{H.}~\bibnamefont{Ebert}},
  {``}\bibinfo{title}{Linear response Kubo-Bastin formalism with application to
  the anomalous and spin Hall effects: A first-principles approach},{''}
  \bibinfo{journal}{Phys. Rev. B} \textbf{\bibinfo{volume}{92}},
  \bibinfo{pages}{184415} (\bibinfo{year}{2015}).

\bibitem[{\citenamefont{Sahin and Flatte}(2015)}]{Sahin2015}
\bibinfo{author}{\bibfnamefont{C.}~\bibnamefont{Sahin}} \bibnamefont{and}
  \bibinfo{author}{\bibfnamefont{M.~E.} \bibnamefont{Flatte}},
  {``}\bibinfo{title}{Tunable Giant Spin Hall Conductivities in a Strong
  Spin-Orbit Semimetal:
  ${\mathrm{Bi}}_{1\ensuremath{-}x}{\mathrm{Sb}}_{x}$},{''}
  \bibinfo{journal}{Phys. Rev. Lett.} \textbf{\bibinfo{volume}{114}},
  \bibinfo{pages}{107201} (\bibinfo{year}{2015}).

\bibitem[{\citenamefont{Peng et~al.}(2016)\citenamefont{Peng, Yang, Singh,
  Savrasov, and Yu}}]{Peng2016}
\bibinfo{author}{\bibfnamefont{X.}~\bibnamefont{Peng}},
  \bibinfo{author}{\bibfnamefont{Y.}~\bibnamefont{Yang}},
  \bibinfo{author}{\bibfnamefont{R.~R.} \bibnamefont{Singh}},
  \bibinfo{author}{\bibfnamefont{S.~Y.} \bibnamefont{Savrasov}},
  \bibnamefont{and} \bibinfo{author}{\bibfnamefont{D.}~\bibnamefont{Yu}},
  {``}\bibinfo{title}{Spin generation via bulk spin current in
  three-dimensional topological insulators},{''} \bibinfo{journal}{Nature
  Communications} \textbf{\bibinfo{volume}{7}}, \bibinfo{pages}{10878}
  (\bibinfo{year}{2016}).

\bibitem[{\citenamefont{Sinitsyn et~al.}(2006)\citenamefont{Sinitsyn, Hill,
  Min, Sinova, and MacDonald}}]{Sinitsyn2006}
\bibinfo{author}{\bibfnamefont{N.~A.} \bibnamefont{Sinitsyn}},
  \bibinfo{author}{\bibfnamefont{J.~E.} \bibnamefont{Hill}},
  \bibinfo{author}{\bibfnamefont{H.}~\bibnamefont{Min}},
  \bibinfo{author}{\bibfnamefont{J.}~\bibnamefont{Sinova}}, \bibnamefont{and}
  \bibinfo{author}{\bibfnamefont{A.~H.} \bibnamefont{MacDonald}},
  {``}\bibinfo{title}{Charge and Spin Hall Conductivity in Metallic
  Graphene},{''} \bibinfo{journal}{Phys. Rev. Lett.}
  \textbf{\bibinfo{volume}{97}}, \bibinfo{pages}{106804}
  (\bibinfo{year}{2006}).

\bibitem[{\citenamefont{Liu et~al.}(2015)\citenamefont{Liu, Zhu, and
  Zheng}}]{Liu2015}
\bibinfo{author}{\bibfnamefont{Z.}~\bibnamefont{Liu}},
  \bibinfo{author}{\bibfnamefont{M.}~\bibnamefont{Zhu}}, \bibnamefont{and}
  \bibinfo{author}{\bibfnamefont{Y.}~\bibnamefont{Zheng}},
  {``}\bibinfo{title}{Quantum transport properties of graphene in the presence
  of randomly distributed spin-orbit coupling impurities},{''}
  \bibinfo{journal}{Phys. Rev. B} \textbf{\bibinfo{volume}{92}},
  \bibinfo{pages}{245438} (\bibinfo{year}{2015}).

\bibitem[{\citenamefont{Garcia and Rappoport}(2016)}]{Garcia2016}
\bibinfo{author}{\bibfnamefont{J.~H.} \bibnamefont{Garcia}} \bibnamefont{and}
  \bibinfo{author}{\bibfnamefont{T.~G.} \bibnamefont{Rappoport}},
  {``}\bibinfo{title}{Kuboâ-Bastin approach for the spin Hall conductivity of
  decorated graphene},{''} \bibinfo{journal}{2D Materials}
  \textbf{\bibinfo{volume}{3}}, \bibinfo{pages}{024007} (\bibinfo{year}{2016}).

\bibitem[{\citenamefont{Sun et~al.}(2016)\citenamefont{Sun, Zhang, Felser, and
  Yan}}]{Sun2016}
\bibinfo{author}{\bibfnamefont{Y.}~\bibnamefont{Sun}},
  \bibinfo{author}{\bibfnamefont{Y.}~\bibnamefont{Zhang}},
  \bibinfo{author}{\bibfnamefont{C.}~\bibnamefont{Felser}}, \bibnamefont{and}
  \bibinfo{author}{\bibfnamefont{B.}~\bibnamefont{Yan}},
  {``}\bibinfo{title}{Strong Intrinsic Spin Hall Effect in the TaAs Family of
  Weyl Semimetals},{''} \bibinfo{journal}{Phys. Rev. Lett.}
  \textbf{\bibinfo{volume}{117}}, \bibinfo{pages}{146403}
  (\bibinfo{year}{2016}).

\bibitem[{\citenamefont{Liu et~al.}(2010)\citenamefont{Liu, Qi, Zhang, Dai,
  Fang, and Zhang}}]{PhysRevB.82.045122}
\bibinfo{author}{\bibfnamefont{C.-X.} \bibnamefont{Liu}},
  \bibinfo{author}{\bibfnamefont{X.-L.} \bibnamefont{Qi}},
  \bibinfo{author}{\bibfnamefont{H.}~\bibnamefont{Zhang}},
  \bibinfo{author}{\bibfnamefont{X.}~\bibnamefont{Dai}},
  \bibinfo{author}{\bibfnamefont{Z.}~\bibnamefont{Fang}}, \bibnamefont{and}
  \bibinfo{author}{\bibfnamefont{S.-C.} \bibnamefont{Zhang}},
  {``}\bibinfo{title}{Model Hamiltonian for topological insulators},{''}
  \bibinfo{journal}{Phys. Rev. B} \textbf{\bibinfo{volume}{82}},
  \bibinfo{pages}{045122} (\bibinfo{year}{2010}).

\bibitem[{\citenamefont{Sinitsyn et~al.}(2004)\citenamefont{Sinitsyn,
  Hankiewicz, Teizer, and Sinova}}]{Sinitsyn2004}
\bibinfo{author}{\bibfnamefont{N.~A.} \bibnamefont{Sinitsyn}},
  \bibinfo{author}{\bibfnamefont{E.~M.} \bibnamefont{Hankiewicz}},
  \bibinfo{author}{\bibfnamefont{W.}~\bibnamefont{Teizer}}, \bibnamefont{and}
  \bibinfo{author}{\bibfnamefont{J.}~\bibnamefont{Sinova}},
  {``}\bibinfo{title}{Spin Hall and spin-diagonal conductivity in the presence
  of Rashba and Dresselhaus spin-orbit coupling},{''} \bibinfo{journal}{Phys.
  Rev. B} \textbf{\bibinfo{volume}{70}}, \bibinfo{pages}{081312}
  (\bibinfo{year}{2004}).

\bibitem[{\citenamefont{Shon and Ando}(1998)}]{Shon1998}
\bibinfo{author}{\bibfnamefont{N.}~\bibnamefont{Shon}} \bibnamefont{and}
  \bibinfo{author}{\bibfnamefont{T.}~\bibnamefont{Ando}},
  {``}\bibinfo{title}{Quantum Transport in Two-Dimensional Graphite
  System},{''} \bibinfo{journal}{J. Phys. Soc. Jpn.}
  \textbf{\bibinfo{volume}{67}}, \bibinfo{pages}{2421} (\bibinfo{year}{1998}).

\bibitem[{\citenamefont{Kato et~al.}(2007)\citenamefont{Kato, Ishikawa, Itoh,
  and ichiro Inoue}}]{Kato2007}
\bibinfo{author}{\bibfnamefont{T.}~\bibnamefont{Kato}},
  \bibinfo{author}{\bibfnamefont{Y.}~\bibnamefont{Ishikawa}},
  \bibinfo{author}{\bibfnamefont{H.}~\bibnamefont{Itoh}}, \bibnamefont{and}
  \bibinfo{author}{\bibfnamefont{J.}~\bibnamefont{ichiro Inoue}},
  {``}\bibinfo{title}{Anomalous Hall effect in spin-polarized two-dimensional
  electron gases with Rashba spin–orbit interaction},{''}
  \bibinfo{journal}{New Journal of Physics} \textbf{\bibinfo{volume}{9}},
  \bibinfo{pages}{350} (\bibinfo{year}{2007}).

\bibitem[{\citenamefont{Hu et~al.}(2008)\citenamefont{Hu, Hwang, and
  Das~Sarma}}]{Hu2008}
\bibinfo{author}{\bibfnamefont{B.~Y.-K.} \bibnamefont{Hu}},
  \bibinfo{author}{\bibfnamefont{E.~H.} \bibnamefont{Hwang}}, \bibnamefont{and}
  \bibinfo{author}{\bibfnamefont{S.}~\bibnamefont{Das~Sarma}},
  {``}\bibinfo{title}{Density of states of disordered graphene},{''}
  \bibinfo{journal}{Phys. Rev. B} \textbf{\bibinfo{volume}{78}},
  \bibinfo{pages}{165411} (\bibinfo{year}{2008}).

\bibitem[{\citenamefont{Chiba et~al.}(2017)\citenamefont{Chiba, Takahashi, and
  Bauer}}]{Chiba2017}
\bibinfo{author}{\bibfnamefont{T.}~\bibnamefont{Chiba}},
  \bibinfo{author}{\bibfnamefont{S.}~\bibnamefont{Takahashi}},
  \bibnamefont{and} \bibinfo{author}{\bibfnamefont{G.~E.~W.}
  \bibnamefont{Bauer}}, {``}\bibinfo{title}{Magnetic-proximity-induced
  magnetoresistance on topological insulators},{''} \bibinfo{journal}{Phys.
  Rev. B} \textbf{\bibinfo{volume}{95}}, \bibinfo{pages}{094428}
  (\bibinfo{year}{2017}).

\bibitem[{\citenamefont{Adroguer et~al.}(2012)\citenamefont{Adroguer,
  Carpentier, Cayssol, and Orignac}}]{Adroguer2012}
\bibinfo{author}{\bibfnamefont{P.}~\bibnamefont{Adroguer}},
  \bibinfo{author}{\bibfnamefont{D.}~\bibnamefont{Carpentier}},
  \bibinfo{author}{\bibfnamefont{J.}~\bibnamefont{Cayssol}}, \bibnamefont{and}
  \bibinfo{author}{\bibfnamefont{E.}~\bibnamefont{Orignac}},
  {``}\bibinfo{title}{Diffusion at the surface of topological insulators},{''}
  \bibinfo{journal}{New Journal of Physics} \textbf{\bibinfo{volume}{14}},
  \bibinfo{pages}{103027} (\bibinfo{year}{2012}).

\bibitem[{\citenamefont{Chen et~al.}(2013)\citenamefont{Chen, Xie, Feng, Yi,
  Liang, He, Mou, He, Peng, Liu et~al.}}]{Chen2013}
\bibinfo{author}{\bibfnamefont{C.}~\bibnamefont{Chen}},
  \bibinfo{author}{\bibfnamefont{Z.}~\bibnamefont{Xie}},
  \bibinfo{author}{\bibfnamefont{Y.}~\bibnamefont{Feng}},
  \bibinfo{author}{\bibfnamefont{H.}~\bibnamefont{Yi}},
  \bibinfo{author}{\bibfnamefont{A.}~\bibnamefont{Liang}},
  \bibinfo{author}{\bibfnamefont{S.}~\bibnamefont{He}},
  \bibinfo{author}{\bibfnamefont{D.}~\bibnamefont{Mou}},
  \bibinfo{author}{\bibfnamefont{J.}~\bibnamefont{He}},
  \bibinfo{author}{\bibfnamefont{Y.}~\bibnamefont{Peng}},
  \bibinfo{author}{\bibfnamefont{X.}~\bibnamefont{Liu}}, \bibnamefont{et~al.},
  {``}\bibinfo{title}{Tunable Dirac Fermion Dynamics in Topological
  Insulators},{''} \bibinfo{journal}{Scientific Reports}
  \textbf{\bibinfo{volume}{3}}, \bibinfo{pages}{2411} (\bibinfo{year}{2013}).

\bibitem[{\citenamefont{Cheng et~al.}(2010)\citenamefont{Cheng, Song, Zhang,
  Zhang, Wang, Jia, Wang, Wang, Zhu, Chen et~al.}}]{Cheng2010}
\bibinfo{author}{\bibfnamefont{P.}~\bibnamefont{Cheng}},
  \bibinfo{author}{\bibfnamefont{C.}~\bibnamefont{Song}},
  \bibinfo{author}{\bibfnamefont{T.}~\bibnamefont{Zhang}},
  \bibinfo{author}{\bibfnamefont{Y.}~\bibnamefont{Zhang}},
  \bibinfo{author}{\bibfnamefont{Y.}~\bibnamefont{Wang}},
  \bibinfo{author}{\bibfnamefont{J.-F.} \bibnamefont{Jia}},
  \bibinfo{author}{\bibfnamefont{J.}~\bibnamefont{Wang}},
  \bibinfo{author}{\bibfnamefont{Y.}~\bibnamefont{Wang}},
  \bibinfo{author}{\bibfnamefont{B.-F.} \bibnamefont{Zhu}},
  \bibinfo{author}{\bibfnamefont{X.}~\bibnamefont{Chen}}, \bibnamefont{et~al.},
  {``}\bibinfo{title}{Landau Quantization of Topological Surface States in
  ${\mathrm{Bi}}_{2}{\mathrm{Se}}_{3}$},{''} \bibinfo{journal}{Phys. Rev.
  Lett.} \textbf{\bibinfo{volume}{105}}, \bibinfo{pages}{076801}
  (\bibinfo{year}{2010}).

\bibitem[{\citenamefont{Zhang et~al.}(2009)\citenamefont{Zhang, Liu, Qi, Dai,
  Fang, and Zhang}}]{Zhang2009}
\bibinfo{author}{\bibfnamefont{H.}~\bibnamefont{Zhang}},
  \bibinfo{author}{\bibfnamefont{C.-X.} \bibnamefont{Liu}},
  \bibinfo{author}{\bibfnamefont{X.-L.} \bibnamefont{Qi}},
  \bibinfo{author}{\bibfnamefont{X.}~\bibnamefont{Dai}},
  \bibinfo{author}{\bibfnamefont{Z.}~\bibnamefont{Fang}}, \bibnamefont{and}
  \bibinfo{author}{\bibfnamefont{S.-C.} \bibnamefont{Zhang}},
  {``}\bibinfo{title}{Topological insulators in Bi2Se3, Bi2Te3 and Sb2Te3 with
  a single Dirac cone on the surface},{''} \bibinfo{journal}{Nat Phys}
  \textbf{\bibinfo{volume}{5}}, \bibinfo{pages}{438} (\bibinfo{year}{2009}).

\bibitem[{\citenamefont{King et~al.}(2011)\citenamefont{King, Hatch, Bianchi,
  Ovsyannikov, Lupulescu, Landolt, Slomski, Dil, Guan, Mi et~al.}}]{King2011}
\bibinfo{author}{\bibfnamefont{P.~D.~C.} \bibnamefont{King}},
  \bibinfo{author}{\bibfnamefont{R.~C.} \bibnamefont{Hatch}},
  \bibinfo{author}{\bibfnamefont{M.}~\bibnamefont{Bianchi}},
  \bibinfo{author}{\bibfnamefont{R.}~\bibnamefont{Ovsyannikov}},
  \bibinfo{author}{\bibfnamefont{C.}~\bibnamefont{Lupulescu}},
  \bibinfo{author}{\bibfnamefont{G.}~\bibnamefont{Landolt}},
  \bibinfo{author}{\bibfnamefont{B.}~\bibnamefont{Slomski}},
  \bibinfo{author}{\bibfnamefont{J.~H.} \bibnamefont{Dil}},
  \bibinfo{author}{\bibfnamefont{D.}~\bibnamefont{Guan}},
  \bibinfo{author}{\bibfnamefont{J.~L.} \bibnamefont{Mi}},
  \bibnamefont{et~al.}, {``}\bibinfo{title}{Large Tunable Rashba Spin Splitting
  of a Two-Dimensional Electron Gas in
  ${\mathrm{Bi}}_{2}{\mathrm{Se}}_{3}$},{''} \bibinfo{journal}{Phys. Rev.
  Lett.} \textbf{\bibinfo{volume}{107}}, \bibinfo{pages}{096802}
  (\bibinfo{year}{2011}).

\bibitem[{\citenamefont{Piot et~al.}(2016)\citenamefont{Piot, Desrat, Maude,
  Orlita, Potemski, Martinez, and Hor}}]{Piot2016}
\bibinfo{author}{\bibfnamefont{B.~A.} \bibnamefont{Piot}},
  \bibinfo{author}{\bibfnamefont{W.}~\bibnamefont{Desrat}},
  \bibinfo{author}{\bibfnamefont{D.~K.} \bibnamefont{Maude}},
  \bibinfo{author}{\bibfnamefont{M.}~\bibnamefont{Orlita}},
  \bibinfo{author}{\bibfnamefont{M.}~\bibnamefont{Potemski}},
  \bibinfo{author}{\bibfnamefont{G.}~\bibnamefont{Martinez}}, \bibnamefont{and}
  \bibinfo{author}{\bibfnamefont{Y.~S.} \bibnamefont{Hor}},
  {``}\bibinfo{title}{Hole Fermi surface in
  ${\mathrm{Bi}}_{2}{\mathrm{Se}}_{3}$ probed by quantum oscillations},{''}
  \bibinfo{journal}{Phys. Rev. B} \textbf{\bibinfo{volume}{93}},
  \bibinfo{pages}{155206} (\bibinfo{year}{2016}).

\bibitem[{\citenamefont{Nomura et~al.}(2014)\citenamefont{Nomura, Souma,
  Takayama, Sato, Takahashi, Eto, Segawa, and Ando}}]{Nomura2014}
\bibinfo{author}{\bibfnamefont{M.}~\bibnamefont{Nomura}},
  \bibinfo{author}{\bibfnamefont{S.}~\bibnamefont{Souma}},
  \bibinfo{author}{\bibfnamefont{A.}~\bibnamefont{Takayama}},
  \bibinfo{author}{\bibfnamefont{T.}~\bibnamefont{Sato}},
  \bibinfo{author}{\bibfnamefont{T.}~\bibnamefont{Takahashi}},
  \bibinfo{author}{\bibfnamefont{K.}~\bibnamefont{Eto}},
  \bibinfo{author}{\bibfnamefont{K.}~\bibnamefont{Segawa}}, \bibnamefont{and}
  \bibinfo{author}{\bibfnamefont{Y.}~\bibnamefont{Ando}},
  {``}\bibinfo{title}{Relationship between Fermi surface warping and
  out-of-plane spin polarization in topological insulators: A view from spin-
  and angle-resolved photoemission},{''} \bibinfo{journal}{Phys. Rev. B}
  \textbf{\bibinfo{volume}{89}}, \bibinfo{pages}{045134}
  (\bibinfo{year}{2014}).

\bibitem[{\citenamefont{Rakyta et~al.}(2012)\citenamefont{Rakyta, Pályi, and
  Cserti}}]{Rakyta2012}
\bibinfo{author}{\bibfnamefont{P.}~\bibnamefont{Rakyta}},
  \bibinfo{author}{\bibfnamefont{A.}~\bibnamefont{Pályi}}, \bibnamefont{and}
  \bibinfo{author}{\bibfnamefont{J.}~\bibnamefont{Cserti}},
  {``}\bibinfo{title}{Electronic standing waves on the surface of the
  topological insulator Bi${}_{2}$Te${}_{3}$},{''} \bibinfo{journal}{Phys. Rev.
  B} \textbf{\bibinfo{volume}{86}}, \bibinfo{pages}{085456}
  (\bibinfo{year}{2012}).

\bibitem[{\citenamefont{{Huynh Duy Khang} et~al.}(2017)\citenamefont{{Huynh Duy
  Khang}, {Ueda}, and {Hai}}}]{2017arXiv170907684H}
\bibinfo{author}{\bibfnamefont{N.}~\bibnamefont{{Huynh Duy Khang}}},
  \bibinfo{author}{\bibfnamefont{Y.}~\bibnamefont{{Ueda}}}, \bibnamefont{and}
  \bibinfo{author}{\bibfnamefont{P.~N.} \bibnamefont{{Hai}}},
  {``}\bibinfo{title}{{A conductive topological insulator with colossal spin
  Hall effect for ultra-low power spin-orbit-torque switching}},{''}
  \bibinfo{journal}{ArXiv e-prints}  (\bibinfo{year}{2017}),
  \eprint{1709.07684}.

\end{thebibliography}

\end{document}